\documentclass[reqno]{article}
\usepackage{amssymb,amsmath,delarray,epsfig,hhline}
\input epsf.tex
\def\DESepsf(#1 width #2){\epsfxsize=#2 \epsfbox{#1}}
\begin{document}

\thispagestyle{empty}

\begin{flushright}
OSU-HEP-02-03\\
UMDPP-02-040
\end{flushright}

\vspace*{0.2in}

\begin{center}\LARGE \bf Neutrino Counting, NuTeV Measurements, Higgs Mass and
$V_{us}$ as Probes of Vectorlike Families in ESSM/SO(10)\end{center}
\vspace*{0.25in}

\centerline{\large\bf K.S. Babu$^{(a)}$ and Jogesh C. Pati$^{(b)}$}

\begin{center}
{(a) Department of Physics, Oklahoma State University\\
Stillwater, OK 74078, USA \\}

{(b) Department of Physics, University of Maryland\\
College Park, MD 20742, USA}

\end{center}






\vskip.50in

\begin{abstract}

\vspace*{0.1in}

The Extended Supersymmetric Standard Model (ESSM), motivated on several
grounds, introduces two vector-like families [${\bf 16}+
{\overline{\bf 16}}$ of SO(10)] with masses of order one TeV.
In an earlier work, a successful pattern
for fermion masses and mixings (to be called pattern I) has been proposed
within a unified SO(10)-framework, based on MSSM,
which makes seven predictions, in good accord with
observations, including $V_{cb}\approx0.04$, and
$\sin^22\theta_{\nu_\mu\nu_\tau}\approx 1$.
Extension of this framework to ESSM, preserving the successes of patern I, has
been recently proposed in an accompanying paper, where it is noted that ESSM
can provide a simple explanation of the indicated anomaly in $(g-2)_\mu$.
To exibit new phenomenological possibilities which may arise within ESSM, we
present here a variant pattern (to be called pattern II) for fermion masses and
mixings, within the SO(10)/ESSM framework, which possesses the same degree of
success as pattern I as regards the masses and mixings of all fermions
including neutrinos. The main point of this paper is to first note that either
one of these two patterns, embedded in ESSM, would lead to a reduction in the
LEP neutrino-counting from $N_\nu=3$ (in good agreement with the data) and also
provide a simple explanation of the $(g-2)_\mu$-anomaly, as pointed out in the
accompanying paper. They can, however, be distinguished from each other by
(a) a sharpening of our understanding of the true magnitude of the
anomaly in $\nu_\mu$-nucleon scattering, recently reported by the NuTeV group,
(b) improved measurements of $m_t$, $m_H$ and $m_W$, (c) improved tests of
$e$-$\mu$ lepton-universality in charged current processes, and (d) improvements
in the measurements of $V_{ud}$ and $V_{us}$. Pattern II (extended to ESSM)
would predict departures from the standard model in the right direction with
regard to (a) and (b), though not as regards (c) and (d) (as judged by the
current data). Pattern I on the other hand would practically coincide with the
standard model as regards its predictions for all four features: (a)-(d). Both
patterns would predict some departure from the SM as regards tau lifetime. The
probes listed above, and, of course, direct searches for the vectorlike families
at the LHC and a future NLC can clearly test ESSM, and even distinguish between
certain variants.
\end{abstract}


\newpage
{\large
\section{Introduction}

The recently reported NuTeV result on
$\nu_\mu$-nucleon scattering \cite{NuTeV} suggests that quite possibly
there is an anomaly in ($\sigma_{NC}/\sigma_{CC}$)-ratios ($R_\nu$ and
$R_{\bar{\nu}}$) compared to expectations of
the Standard Model. The results on $R_\nu$ and $R_{\bar{\nu}}$ have been
interpreted in
Ref. \cite{NuTeV} to reflect either (a) a higher on-shell value of
$\sin^2\theta_W$ which is at $3\sigma$ above the value extracted from
other experiments within the
Standard Model (SM), or (b) a reduced coupling of the left-handed quarks to
$Z^0(g_L^{\rm eff})$, compared to the SM value for the same. A third
possibility
has also been mentioned in the context of a two-parameter fit corresponding
to a reduced overall strength ($\rho_0$) of the neutral current four
fermion coupling together with a possible non-standard value of
$\sin^2\theta_W$.
If the result persists against (even) more precise data,
and improved theoretical scrutiny involving Standard Model effects,
it would clearly have some profound implications.

In addition to the NuTeV anomaly [counted as (a)], there also exist a few other
possible discrepancies in the predictions of the Standard Model (SM). These in
particular include: (b) the LEP neutrino-counting \cite{27} which is about
$2\sigma$ below the SM value of 3, (c) the empirical Higgs mass limit
$m_H>115$ GeV \cite{lephiggs} (while typically the SM would suggest
$m_H<100$ GeV, see later), (d) the measured value of $m_W=80.446\pm 0.040$ GeV
(LEP data), which is typically higher than the SM value (see discussions
below), and (e) the possible 1.6 to 3$\sigma$ discrepancy in $(g-2)_\mu$
(for experimental and theoretical analysis, see respectively Ref. \cite{BNL}
and \cite{theorg-2}) \cite{6}. The purpose of this note is to seek for
{\it correlated explanations} of the two indicated discrepancies
(a) and (b), together with desired shifts in $m_H$ and $m_W$,
in the context of an old idea motivated on several grounds (see
below) \cite{JCP_BPS_BPZ,BP}. A simple explanation of the indicated anomaly
in $(g-2)_\mu$, together with associated tests through radiative processes
($\tau\rightarrow \mu\gamma$ and $\mu\rightarrow e\gamma$), in the context of
the same idea, has been noted in an accompanying paper \cite{BP(g-2)}.

The idea in question pertains to the so-called ``Extended SuperSymmetric
Standard Model" (ESSM), which introduces two complete vectorlike families
of quarks and leptons -- denoted by  $Q_{L,R}=(U,D,N,E)_{L,R}$ and
$Q'_{L,R}=(U',D',N',E')_{L,R}$ -- with masses of order few hundred GeV to
one TeV. Both $Q_{L}$ {\it and} $Q_{R}$ transform as (2,1,4), while
$Q'_{L}$ {\it and} $Q'_{R}$ transform as (1,2,4) of the quark-lepton
unifying symmetry G(224)=SU(2)$_L\times$SU(2)$_R\times$SU(4)$^C$.
Thus, together, they transform as a pair {\bf 16}+${\bf \overline{16}}$ of
SO(10), to be denoted by ${\bf 16_V}=(Q_L|\bar{Q}'_R)$ and
${\bf \overline{16}_V}=(\bar{Q}_R|Q'_L)$. The subscript ``V" signifies two
features: (a) ${\bf \overline{16}_V}$ combines primarily with ${\bf 16_V}$,
 so that the pair gets an SO(10)-invariant
(thus SU(2)$_L\times$U(1)-invariant) mass-term of the form
$M_V{\bf 16_V\cdot\overline{16}_V}+
h.c.=M_V(\bar{Q}_RQ_L+\bar{Q}'_RQ'_L)+h.c.$, at the GUT scale,
utilizing for example the VEV of an SO(10)-singlet, where $M_V\sim$ few
hundred GeV to one TeV \cite{FN4}, (b) since $Q_L$ and $Q_R$ are
doublets of SU(2)$_L$, the massive four-component object
$(Q_L\oplus Q_R)$ couples {\it vectorially} to $W_L$'s; likewise
$(Q'_L\oplus Q'_R)$ couples {\it vectorially} to $W_R$'s. Hence the name
"$\underline{\mbox{vectorlike}}$" families.

 It has been observed in earlier
works \cite{BPzhang} that addition of complete {\it vectorlike}
families [{\bf 16}+${\bf \overline{16}}$ of SO(10)], with masses
$\gtrsim 200\;$ GeV to one TeV (say), to the Standard Model naturally
satisfies all the phenomenological constraints so far. These include:
(a) neutrino-counting at LEP 
(because $M_{N,N'}>m_Z/2$),
(b) measurement of the $\rho$-parameter [because the SO(10)-invariant mass
for
the vectorlike families ensure up-down degeneracy -- i.e., $M_U=M_D$, etc. --
to a good accuracy], and (c) those of the oblique electroweak parameters
\cite{Altarelli_PeskinTakeuchi} (for the same reasons as indicated above)
\cite{ParticleData}. We will comment in just a moment on the theoretical
motivations for ESSM. First let us note why ESSM is expected to be relevant
to the NuTeV anomaly and why it would simultaneously have implications for
the LEP neutrino-counting. As a central feature, ESSM assumes that the three
 chiral
families ($e$, $\mu$ and $\tau$) receive their masses primarily (barring
corrections $\lesssim$ a few MeV) through their mixings with the two
vectorlike families \cite{JCP_BPS_BPZ,BP}.
As we will explain in Sec. 2, this feature has the advantage that it
automatically renders the electron family massless (barring corrections as
mentioned above); and at the same time it naturally assigns a large hierarchy
between the muon and the tau family masses, {\it without} putting in such a
hierarchy in the respective Yukawa couplings \cite{JCP_BPS_BPZ,BP,FN8}. In
short, ESSM provides a simple reason for the otherwise mysterious interfamily
mass hierarchy, i.e., ($m_{u,d,e}\ll m_{c,s,\mu}\ll m_{t,b,\tau}$).
Now, since the chiral families get masses by mixing with the vectorlike
families, the observed neutrinos $\nu_i$ naturally mix with the heavy
neutrinos $N_L$ and $N'_L$ belonging to the families $Q_L$ and $Q'_L$,
respectively. The mixing parameters get determined in terms of fermion
masses and mixings. As we will explain, it is the mixing of $\nu_\mu$ and
likewise of $\nu_\tau$ with the SU(2)$_L$-{\it singlet} heavy lepton $N'$
belonging to the family $Q'_{L}$, that reduces the overall strengths of the
couplings (i) $Z^0 \rightarrow \nu_\mu \bar{\nu}_\mu$,
(ii) $Z^0 \rightarrow \nu_\tau \bar{\nu}_\tau$, as well as of
(iii) $W^+\rightarrow \mu^+ \nu_\mu$, and (iv) $W^+\rightarrow \tau^+\nu_\tau$,
compared to those of the Standard Model, {\it all in a predictably correlated
manner}. The forms of the couplings remain, however, the same as in the
Standard Model.

In accord with the interfamily hierarchy of fermion masses and mixings, the
reduction in the couplings as above is found to be family-dependent, being
maximum in the $\nu_\tau$, intermediate in $\nu_\mu$ and small ($<$ one
part in $10^4$) in the $\nu_e$-channel.

These effects would manifest themselves as (a) a deficit in the LEP
neutrino-counting from the Standard Model value of $N_\nu\equiv N_{\nu_e}+
N_{\nu_\mu}+N_{\nu_\tau}=3$, (b) as a correlated reduction in the
strength of $\nu_\mu N\rightarrow\nu X$ interaction (which is relevant
to the NuTeV anomaly), as (c) small departures from lepton universality in the
charged current processes, and as (d) a decrease in $V_{ud}$ (depending upon
$u$-$U'$ mixing) compared to its
SM value. Qualitative aspects of these effects arising
from $\nu_i$-$N'$-mixing (without a quantitative hold on the reduction in the
$\nu_\mu \bar{\nu}_\mu$ and
$\nu_\tau \bar{\nu}_\tau$--couplings to $Z^0$) were in fact noted in an earlier
 work
\cite{BPzhang} almost ten years ago. In that work, motivated by an
(overly) simplified version of understanding the inter-family hierarchy, the
effect of the $\nu_\mu \bar{\nu}_\mu$-channel was considered to be too small.
Two interesting developments have, however, taken place in the meanwhile.
First, SuperK discovered atmospheric neutrino oscillations, showing that
$\nu_\mu$ oscillates very likely into $\nu_\tau$ with a surprisingly large
oscillation angle: $\sin^22\theta^{\rm osc}_{\nu_\mu\nu_\tau}\gtrsim 0.92$
\cite{SuperKatm}. Second, motivated in part by the SuperK result, an
economical SO(10)-framework, based on MSSM, has been proposed in the context
of a minimal
Higgs system (${\bf 10_H}$, ${\bf 16_H}$, ${\bf \overline{16}_H}$ and
${\bf 45_H}$) to address the problem of fermion masses and mixings \cite{BPW}.
 Within
this framework, a few variant patterns of fermion mass-matrices are possible,
each of which is extremely successful in describing the masses and mixings
of all
fermions including neutrinos. For example, the pattern exhibited
in \cite{BPW} (to be called pattern I) makes seven predictions,
including $V_{cb}\approx 0.042$ and
$\sin^22\theta^{\rm osc}_{\nu_\mu\nu_\tau}\approx 0.85$-0.99, all in accord
with the data to within 10\% \cite{other}. Extension of this SO(10)-framework
to ESSM, preserving the successes of pattern I, has recently been proposed in
an accompanying paper \cite{BP(g-2)}, where it is noted that ESSM can provide a
simple explanation of the indicated anomaly in $(g-2)_\mu$ without requiring
the presence of light sleptons.

To exibit new phenomenological possibilities, which may asrise in ESSM, we
present here a {\it variant pattern (to be called pattern II)} for fermion
masses and mixings, within the SO(10)-framework, incorporating ESSM. This
variant possesses the same degree of success as pattern I as regards the masses
and mixings of all fermions including neutrinos. One of the main points of this paper
is to  note that both patterns (I or II), embedded in ESSM, would lead to
a reduction in LEP neutrino-counting from $N_\nu=3$, owing to a mixing of
$\nu_\mu$ and $\nu_\tau$ with the singlet heavy neutrino $N'$ (in good
agreement with the data), and would also provide a simple explanation of the
indicated anomaly in $(g-2)_\mu$, as pointed out in the accompanying paper
\cite{BP(g-2)}.

Interestingly enough, we observe that the two variants (I and II) can clearly
be distinguished from each other, however, by other phenomena. In particular,
owing to a relative enhancement of $\nu_\mu-N'$ mixing, pattern II, to be
presented in sec. 3, can provide (a) a partial explanation of the anomaly in
$\nu_\mu$-nucleon scattering reported recently by the NuTeV group, and
simultaneously (b) an increase in the predictions for $M_H$ and $m_W$ compared
to those of the standard model (in good accord with experiments), (c) small
departures from $e$-$\mu$ universality (up to 1 to 1.8$\sigma$ deviation from
experiments), and (d) small decreases in the effective values of $V_{ud}$ and
$V_{us}$ compared to those in the standard model, which are currently
disfavored by the data on grounds of unitarity of the $3\times3$ CKM-matrix
(this latter turns out to be well respected even for ESSM, especially for the
sum of the absolute squares of the entries in the first row). Patern I, proposed
in \cite{BPW}, embedded in ESSM \cite{BP(g-2)}, on the other hand, practically
coincides with the standard model as regards its predictions for all four
features: (a)-(d). Both patterns (I and II) would lead to some
departures from the standard model (though of differing magnitudes) as regards
their predictions for
the tau lifetime. Improved experimental and theoretical studies involving these
five features can thus clearly distinguish between the two variants (I and II)
and thereby shed light on GUT/string-scale physics.

We stress that the two variants I and II, embedded in ESSM, while differing
as regards (a)-(d), share the common feature that they both depart from the
standard model as regards their predictions for $N_\nu$ and $(g-2)_\mu$, in good
accord with the present data.

Before discussing the relevance of ESSM to the phenomena mentioned
above, a few words about motivations for ESSM
 might be in order. Note that it, of course, preserves all the merits of
 MSSM as regards gauge coupling unification and protection of the Higgs
 masses against large quantum corrections. Theoretical motivations for the
 case of ESSM arise on several grounds: (a) It provides a better chance for
 stabilizing the dilaton by having a semi-perturbative value for
 $\alpha_{\rm unif}\approx 0.25$ to 0.3 \cite{BP}, in contrast to a very weak
 value of 0.04 for MSSM; (b) It raises the unification scale $M_X$
 \cite{BP, HemplingKolda_Mar} compared to that for MSSM and thereby reduces
 substantially the mismatch between MSSM and string unification scales
 \cite{Ginspang_Kaplunovsky_Dienes}; (c) It lowers the GUT-prediction
 for $\alpha_3(m_Z)$ compared to that for MSSM \cite{BP}, as needed by the
 data; (d) Because of (b) and (c), it naturally enhances the GUT-prediction
 for proton lifetime compared to that for MSSM embedded in a GUT
 \cite{BPW,JCP_Erice}, also as needed by the data
 \cite{SKlimit_protonlifetime}; and finally, (e) as mentioned above, it
 provides a simple reason for interfamily mass-hierarchy. In this sense,
 ESSM, though less economical than MSSM, offers some distinct advantages.

In an accompanying paper \cite{BP(g-2)} we have noted that ESSM provides a
simple explanation of the indicated anomaly in $(g-2)_\mu$ and that such an
explanation can be clearly tested by improved searches for
$\tau\rightarrow \mu\gamma$ and $\mu\rightarrow e\gamma$-decays. The
 main purpose of this paper is two-fold: (1) First, we present in sec. 3 a
{\it variant pattern} (called pattern II) of fermion masses and mixings within
SO(10)-framework, {\it which is interesting in its own right, in that it has
the same degree of success as pattern I} \cite{BP(g-2),BPW} {\it in describing
fermion masses and mixings, regardless of whether it is embedded in ESSM or
not.} (2) Second, we study (in secs. 4, 5 and 6) the phenomenological
consequences
of pattern II, embedded in ESSM and those for pattern I in sec. 7. We observe
that pattern II (thus embedded) not
only leads to a decrease in LEP neutrino counting from $N_\nu=3$ (like pattern
I), but also leads to departures from standard model predictions as regards
(i) NuTeV-type measurements, (ii) masses of the Higgs and W-bosons, (iii)
lepton-universality and (iv) effective values of $V_{ud}$ and $V_{us}$ (unlike
pattern I). These issues are discussed in secs. 4, 5 and 6 and a comparison
between
the predictions of pattern I and pattern II is made in sec. 7. A summary is
presented in sec. 8.

\section{Fermion Masses and Mixings in ESSM}

Following the discussion in the introduction (see Ref. \cite{BP} for details
and notation), the 5$\times$5 mass-matrix involving the three chiral
$(q^i_{L,R})$ and the two vectorlike families ($Q_{L,R}$ and $Q'_{L,R}$)
is assumed to have the see-saw form:
\begin{eqnarray}
\label{eq:MYuk}
\begin{array}{cc}
& \begin{array}{lcr}  q^i_L\;\;\;\; &\;\;\;\; Q_L & \;\;\;\;\;\;Q'_L
\end{array}\\
M^{(0)}_{f,c}=\begin{array}{c} \bar{q}^i_R\\
                                 \bar{Q}_R\\
                                 \bar{Q}'_R
\end{array} &
\begin{array}({ccc}) 0_{3\times 3} & X_f\langle H_f\rangle
& Y_c\langle H_s\rangle \\
                    Y^{\prime \dagger}_c\langle H_s\rangle
& z_c\langle H_V \rangle& 0\\
                    X^{\prime \dagger}_f\langle H_f\rangle  & 0
& z'_f\langle H_V\rangle
\end{array}
\end{array}~.
\end{eqnarray}
Here the symbols $q$, $Q$ and $Q'$ stand for quarks as well as leptons;
$i$=1, 2, 3 corresponds to the three chiral families. The subscript $f$
for the Yukawa-coupling column matrices $X_f$ and $X'_f$ denotes $u$, $d$,
$l$ or $\nu$, while $c=q$ or $l$ denotes quark or lepton color. The fields
$H_f$ with $f=u$ or $d$ denote the familiar two Higgs doublets, while
$H_s$ and $H_V$ are Higgs Standard Model singlets \cite{FN17}, whose
VEVs are as follows: $\langle H_V \rangle\equiv v_0\sim 1\;$TeV $ \gtrsim
\langle H_s\rangle\equiv v_s \gtrsim \langle H_u \rangle \equiv v_u
\sim 200\; $GeV$~ \gg\langle H_d \rangle\equiv v_d$. The zeros in Eq.
\eqref{eq:MYuk}, especially the
direct coupling terms appearing in the upper 3$\times$3 block, are expected
to be corrected so as to leads to masses $\lesssim$ a few MeV.

The parametrization in Eq. \eqref{eq:MYuk} anticipates that differences
between $z_c$ and $z'_f$, between $X_f$ and $X'_f$, and between $Y_c$ and
$Y'_c$ may arise at the electroweak scale in part because renormalization
effects distinguish between $Q_{L,R}$, which are SU(2)$_L$-doublets, and
$Q'_{L,R}$, which are SU(2)$_L$-singlets (see Eq. (10) of Ref. \cite{BP}),
and in part because ($B$-$L$)-dependent and $L$-$R$ as well as
family-antisymmetric contributions may arise effectively by utilizing the
VEV of a ${\bf 45}$-plet, sometimes in conjunction with that of a ${\bf 10_H}$
(see Refs \cite{BPW} and \cite{BP(g-2)} for details), which can introduce
differences between $X_f$ and $X'_f$, etc.

Denoting $X^T_f=(x_1,x_2,x_3)_f$ and $Y^T_c=(y_1,y_2,y_3)_c$, it is easy
to see \cite{JCP_BPS_BPZ,BP,FN8} that regardless of the values of these Yukawa
couplings, one can always transform the basis vectors $\bar{q}_R^i$ and
$q_L^i$ so that $Y_c^T$ transforms into $\hat{Y}^T_c=(0,0,1)y_c$,
$X^T_f$ simultaneously into $\hat{X}^T_f=(0,p_f,1)x_f$, $X^{\prime \dagger}_f$
into $\hat{X}^{\prime \dagger}_f=(0,p'_f,1)x'_f$ and $Y^{\prime \dagger}_c$
into
$\hat{Y}^{\prime \dagger}_c=(0,0,1)y'_c$. It is thus apparent why one family
remains
massless (barring corrections of $\lesssim$ a few MeV), despite lack of
any hierarchy in the Yukawa couplings $(x_i)_f$ and $(y_i)_c$, etc. This
one is naturally identified with the electron family.
To a good approximation, one also obtains the relations \cite{JCP_BPS_BPZ,BP}:
$m_{c,s,\mu}^0 \approx m^0_{t,b,\tau}(p_fp_f'/4)$.  Even if $p_f,p_f'$
are not so small (e.g. suppose $p_f, p_f' \sim 1/2$ to $1/4$), their product
divided by four can still be pretty small.  One can thus naturally get
a large hierarchy between the masses of the muon and the tau families
as well. We note that while the transfomation of the $X$, $X'$, $Y$ and $Y'$
matrices into their hat-forms as above demonstrates the masslessness of the
electron family, such transformations would in general induce some mixing
among the three chiral families, which could be important for the first family.
Ordinarily we may, however, expect the entries of the $X$ and $X'$ matrices in
the gauge basis to be hierarchical ($x_1\ll x_2\ll x_3$, etc.) owing
to flavor symmetries. In this case, such mixings would be unimportant. We will
assume this to be the case in the text, but allow for possible departures from
this assumption in the appendix.

As shown in Ref. \cite{BP(g-2)}, the SO(10) group-structure of the
(2,3)-sector of the effective 3$\times$3 mass matrix for the three chiral
families, proposed in Ref. \cite{BPW}, can be preserved (to a good
approximation) for the case of ESSM, simply by imposing an SO(10)-structure
on the off-diagonal Yukawa couplings of Eq. \eqref{eq:MYuk}, that is
analogous to that of Ref. \cite{BPW} (see \cite{FN18}), while small entries
involving the first family can be inserted, as in Ref. \cite{BPW}, through
higher dimensional operators. (We refer the reader to Ref. \cite{BP(g-2)}
and to a forthcoming paper \cite{BPforthcoming} devoted entirely to
``fermion masses in ESSM" for more details.)

It is the Dirac mass-matrices of the neutrinos and of the charged leptons
that are relevant to the present paper. In the hat-basis mentioned above,
where the first family is (almost) decoupled from the two vectorlike families,
the Dirac mass-matrix of the neutrinos (following notations of Ref. \cite{BPW}
and \cite{BP(g-2)}) is given by \cite{FN20}:
\begin{eqnarray}
\label{eq:MD}
\begin{array}{cc}
& \begin{array}{ccccc} \nu_L^e\,& \nu_L^\mu \,
& \nu_L^\tau \quad\quad\;\; & N_L \qquad\quad\quad& N'_L \qquad\quad
\end{array}\\
M^{D}_\nu=\begin{array}{c} \bar{\nu}_R^e\\
                           \bar{\nu}_R^\mu\\
                           \bar{\nu}_R^\tau\\
                           \bar{N}_R\\
                           \bar{N}'_R
\end{array} &
\begin{array}({ccc}) 0_{3\times 3} &
\begin{array}({c}) 0\\
                   p_\nu\\
                   1
\end{array}\kappa^\nu_u &
\begin{array}({c}) 0\\
                   0\\
                   1
\end{array}\kappa^l_s\\
\left( 0 \;\;\;\;\; 0 \;\:\;\;\; 1 \right) \kappa ^{\prime l}_s & M_N & 0\\
\left( 0 \;\;\;\; p'_\nu \;\;\;\; 1 \right) \kappa ^{\prime \nu}_u & 0 & M_{N'}
\\
 M_{N'}
\end{array}
\end{array} ~.
\end{eqnarray}
Here, $\kappa_u^\nu\equiv x_\nu\langle H_u\rangle$,
$\kappa^{\prime \nu}_u\equiv x'_\nu\langle H_u\rangle$,
 $\kappa_s^l\equiv y_l\langle H_s\rangle$,
 $\kappa^{\prime \nu}_s\equiv y'_l\langle H_s\rangle$,
 $M_N\approx M_E=z_l\langle H_V\rangle$,
 $M_{N'}\approx M_{E'}=z'_l\langle H_V\rangle$. The mass matrix for the
 charged leptons is obtained by replacing the suffix $\nu$ by $l$ and $u$ by
 $d$, so that $H_u\rightarrow H_d$, $\kappa_u^\nu\rightarrow \kappa_d^l$, but
$\kappa_s^l\rightarrow \kappa_s^l$, etc. Analogous substitution give the mass
 matrices
for the up and down quarks. We stress that {\it the parameters of the
mass-matrices of the four sectors $u$, $d$, $l$ and $\nu$, and also those
entering into $X$ versus $X'$ or $Y$ versus $Y'$ in a given sector, are of
course not all independent,} because a large number of them are related to
each other at the GUT-scale by the group theory of SO(10) and the
representation(s) of the relevant Higgs multiplets \cite{FN21}. For
convenience of writing, we drop the superscript $\nu$ on kappas, from now on.

We now proceed to determine some of these parameters in the context of
a promising SO(10)--model,
which  would turn out to be especially relevant to the NuTeV--anomaly
and the LEP neutrino counting.

\section{A Variant Pattern For Fermion Masses Within the SO(10)/ESSM Framework}

Following the approach of \cite{BPW}
we now present a concrete example wherein the effective mass-matrices for
ESSM, exhibited in Eq. \eqref{eq:MYuk} and \eqref{eq:MD}, emerge from a
unified SO(10) framework, and which turns out to be relevant to the NuTeV
anomaly.
The pattern of the mass-matrices for the three light families in the
($u,d,l,\nu$)-sectors, which result from this example upon integrating out
 the heavy
families ($Q$ and $Q'$), turns out to be a {\it simple
variant} of the corresponding pattern presented in Ref. \cite{BPW}.
The ESSM-extension \cite{BP(g-2)} of the pattern of Ref. \cite{BPW} will be
called Model I, while that of the variant presented here will be called Model
II. The variant
preserves the economy (in parameters) and the successes of Ref. \cite{BPW}
 as regards  predictions of the masses and mixings of quarks as well as
leptons including neutrinos; these include $V_{cb} \simeq 0.04$ and
$\sin^22\theta_{\nu_\mu \nu_\tau}^{\rm osc} \approx 1$.  At the same time,
the variant, extended to ESSM (Model II), turns out to be relevant to
partially account for the
NuTeV-anomaly and simultaneously for the LEP data on neutrino counting, while
Model I \cite{BP(g-2),BPW} can account for the LEP neutrino counting, but not
for the NuTeV anomaly.

Let us first assume for the sake of simplicity that the electron family is
(almost)
decoupled from the heavy families ($Q$ and $Q'$) in the gauge-basis -- that
is to say, the gauge and the hat--basis (defined earlier) are essentially
the same, so that Eq. \eqref{eq:MD} holds to a good approximation,
already in the gauge basis. (In the appendix we will consider possible
departures from this assumption.)
Consider then the following superpotential, which involves the $\mu$ and the
$\tau$
families (${\bf 16}_2$ and ${\bf 16}_3$) and the two vector-like families
(${\bf 16}_V$ and $\overline{\bf 16}_V$):
\begin{eqnarray}
\label{eq:WrmYuk}
W_{\rm Yuk} &=& h_V {\bf 16}_V \overline{\bf 16}_V H_V + h_{3V} {\bf 16}_3
{\bf 16}_V {\bf 10}_H + h_{3V}' {\bf 16}_3{\bf 16}_V{\bf 16}_H {\bf 16}_H^d/M
\nonumber \\
&+& h_{3\bar{V}}{\bf 16}_3 \overline{\bf 16}_VH_s +
\tilde h_{2V} {\bf 16}_2{\bf 16}_V
{\bf 10}_H {\bf 45}_H/M + h_{2V}'{\bf 16}_2{\bf 16}_V{\bf 16}_H^d{\bf 16}_H/M~.
\end{eqnarray}
Contrast this from the superpotential of
Model I given in Ref. \cite{BP(g-2)}.
Here, $\left\langle {\bf 16}_H \right\rangle \sim \left\langle {\bf 45}_H
\right \rangle
\sim \left\langle X \right\rangle\sim M_{\rm GUT}$, and
$M \sim M_{\rm string}$, with
$X$ being an SO(10) singlet and $\left\langle {\bf 45}_H \right
\rangle$ being proportional to $B-L$.  As mentioned before,
$\left\langle H_V \right
\rangle \sim 1$ TeV $>\left\langle H_s \right\rangle\sim\left\langle H_u
\right
\rangle \sim 200$ GeV $\gg \left\langle H_d \right\rangle$. ${\bf 10}_H$
contains
the Higgs doublets $H_u$ and $H_d$ of MSSM. While $H_u = {\bf 10}_H^u$, the
down type Higgs is contained partly in ${\bf 10}_H^d$ and partly in ${\bf 16}
_H^d$
(see Ref. \cite{BPW}) -- that is $H_d=\cos\gamma{\bf 10}_H^d+\sin\gamma
{\bf 16}_H^d$. If
$\sin\gamma = 0$, one would have $\tan\beta = m_t/m_b$, but with
$\cos\gamma \ll 1$, $\tan\beta$ can have small to intermediate values of
$3-20$.
($\tan\beta/\cos\gamma = m_t/m_b$ is fixed.)  The entries in Eq.
\eqref{eq:WrmYuk} with a
factor $1/M$ are suppressed by $M_{\rm GUT}/M \sim 1/10$.  Note however
that the
contribution from $h_{3V}$ and $h_{3V}'/M$ terms to the down quark and charged
lepton mass matrices could be comparable, $\cos\gamma \sim 1/10$, which is
what we adopt.

One can verify that Eq. \eqref{eq:WrmYuk} will induce mass--matrices of the
type shown in
Eqs. \eqref{eq:MYuk} and \eqref{eq:MD} with definite correlations among
$X_f, X_f', Y_c$ and
$Y_c'$ sectors.  To see these correlations, it is useful to
block--diagonalize
the $5 \times 5$ mass matrix given by Eq. \eqref{eq:MD} and its analogs, so
that
the light families ($e, \mu$ and $\tau$) get decoupled from the heavy ones.
From now on, we denote the gauge basis (in which Eqs. (1) and (2) are written)
by $\psi_{L,R}^0$ and the transformed basis which
yields  the block--diagonal form by $\psi_{L,R}'$.  Given the SO(10)
group structure of Eq. \eqref{eq:WrmYuk}, it is easy to see that the
effective Dirac
mass matrices of the muon and the tau families  in the
up, down, charged lepton and neutrino sectors, resulting from
block-diagonalization, would have the following form at the GUT scale
\cite{FN25}, which is referred to as pattern II:
\begin{eqnarray}
\label{eq:4matrices}
\begin{array}{cc}
M_u = \left(\begin{array}{cc} 0 & -\epsilon \\ \epsilon & 1
\end{array}\right){\cal M}_{u}^0, &
M_d = \left(\begin{array}{cc} 0 & \eta-\epsilon \\ \eta+\epsilon & 1+\xi
\end{array}\right){\cal M}_{d}^0 ,
\\
M_\nu^D = \left(\begin{array}{cc} 0 & 3\epsilon \\ -3\epsilon & 1
\end{array} \right){\cal M}_{u}^0, &
M_l = \left(\begin{array}{cc} 0 & \eta+3\epsilon \\ \eta-3\epsilon & 1+\xi
\end{array}\right){\cal M}_{d}^0 .
\end{array}
\end{eqnarray}
Here the matrices are written in the primed basis (see above), so that the
Lagrangian is given by ${\cal L} = \overline{\psi}'_R {\cal M}\psi_L'+h.c.$ .
 (These
should be compared with the transpose of the corresponding matrices in Ref.
\cite{BPW}.) It is easy to verify that the entries $1.{\cal M}_u^0$, $\xi$,
$\epsilon$, $\eta$ and $1.{\cal M}_d^0$ are proportional respectively to
$h_{3V}h_{3\bar{V}}/h_{V}$, $h'_{3V}h_{3\bar{V}}/h_{V}$,
$\tilde h_{2V}h_{3\bar{V}}/h_{V}$, $h'_{2V}h_{3\bar{V}}/h_{V}$ and
$h_{3V}h_{3\bar{V}}/h_{V}$. (For example $({\cal M}_{u}^0,{\cal M}_{d}^0)\simeq
(m_t^0,m_b^0)\simeq (2h_{3V}h_{3\bar{V}}/h_{V})(v_u,v_d)(v_s/v_0)$.) Note that
 the
$(B-L)$-dependent antisymmetric parameter $\epsilon$ arises because
$\langle {\bf 45}_H\rangle\propto B-L$.

The eight $p$-parameters of Eq. \eqref{eq:MD} and its analogs can be readily
obtained
from Eqs. \eqref{eq:WrmYuk} and \eqref{eq:4matrices}.  They are (for the
present case of Model II):
\begin{eqnarray}
\label{eq:8p}
p_u &=& -2\epsilon,~ p_u'=2\epsilon,~ p_d =\frac {2(\eta-\epsilon)} {1+\xi},~
p_d' = \frac{2 (\eta+\epsilon)}{1+\xi},\nonumber \\
p_\nu &=& 6 \epsilon,~ p_\nu' = -6\epsilon,~ p_l = \frac{2(\eta+3\epsilon)
}{1+\xi},~
p_l'=\frac{2(\eta-3\epsilon)}{1+\xi}~.
\end{eqnarray}
As mentioned earlier, there are only three independent parameters
($\eta,\epsilon,\xi$),
leading to nontrivial correlations between observables.

The matrices of Eq. \eqref{eq:4matrices} can be diagonalized in the
approximation $\epsilon,\eta \ll \xi,1$.  One obtains (for pattern II):
\begin{eqnarray}
\label{eq:6}
\frac{m_b^0}{m_\tau^0} &\simeq& 1 - \frac{8 |\epsilon|^2}{|1+\xi|^2},~~~
\frac{m_c^0}{m_t^0} \simeq |\epsilon|^2,~~~ \frac{m_s^0}{m_b^0} \simeq
\frac{|\eta^2-\epsilon^2|}{|1+\xi|^2},\nonumber\\
\frac{m_\mu^0}{m_\tau^0} &\simeq& \frac{|\eta^2-9 \epsilon^2|}{|1+\xi|^2},~
\quad\qquad
|V_{cb}^0| \simeq \frac{|\epsilon \xi - \eta|}{|1+\xi|}~.
\end{eqnarray}
Here the superscript ``0" denotes that these relations hold at the unification
scale.  A reasonably good fit to all observables can be obtained (details of
this discussion will be given in a separate paper \cite{BPforthcoming}) by
choosing
\begin{equation}
\label{eq:values}
\epsilon = -0.05,~ \eta = 0.0886,~ \xi = -1.45,\mbox{ (Model II)}
\end{equation}
which leads to \cite{FN27}
$m_\mu^0/m_\tau^0 \simeq 1/17.5,~ m_s^0/m_b^0 \simeq 1/44.5$, $m_c^0/m_t^0
\simeq 1/400,~ |V_{cb}^0| \simeq 0.031$.  After renormalization group
extrapolation is used (using $\tan\beta = 10$ for definiteness) these values
lead to $m_c(m_c) = 1.27$ GeV, $|V_{cb}| = 0.036,~ m_s(1~{\rm GeV}) = 160$
MeV, all of which are in good agreement with observations (to within 10\%).
Owing to the larger QCD effect in ESSM compared to MSSM, the predicted value
of $m_b(m_b)$ is about 20\% larger than the experimentally preferred value.
Allowance for either larger values of $\tan\beta$ ($\approx 35-40$)
\cite{bagger},
or gluino threshold corrections, and/or a 20\% $B-L$ dependent correction
to the vector family mass at the GUT scale (see Ref. \cite{BP(g-2)}) could
account for such a discrepancy.

Eqs. \eqref{eq:8p} and \eqref{eq:values} lead to \cite{FN26}:
\begin{eqnarray}
\label{eq:8newp}
\left. \begin{array}{cccc}
p_\nu = -0.30,& p_\nu' = +0.30,& p_l = 0.282,& p_l' = -1.05
\\
p_u = 0.10,& p_u' = -0.10,& p_d = -0.607,& p_d' = -0.163~.
\end{array}\right\}(\mbox{Model II})
\end{eqnarray}

Now, the light neutrino masses are induced by the seesaw mechanism.  In
addition to Eq. \eqref{eq:WrmYuk}, there are terms in the superpotential
that induce heavy Majorana masses for the right handed neutrinos of the
three chiral
families: $W'\supset\nu_R^{\prime T} M_R^\nu \nu_R'$. Let us assume that the
matrix $M_R^\nu$ for the dominant $\nu_\mu-\nu_\tau$ sector has the simple
form
\begin{eqnarray}
M_R^\nu = M_R \left(\begin{array}{cc}0 & y \\ y & 1\end{array}\right)
\nonumber
\end{eqnarray}
 as in Ref. \cite{BPW}.
(We are ignoring here the masses and mixings of the first family. Their
inclusion will modify the present discussion only slightly.)  The effective
light neutrino mass matrix for the $\nu_\mu-\nu_\tau$ sector is them
\begin{eqnarray}
\label{eq:Mnu}
M_\nu^{\rm light} = \frac{1}{y^2 M_R} \left(\begin{array}{cc}
0 & y \epsilon^2 \\
y \epsilon^2 & \epsilon^2+2\epsilon y \end{array}\right)~.
\end{eqnarray}
The $\nu_\mu-\nu_\tau$ oscillation angle is then
\begin{equation}
\label{eq:angle}
\theta_{\nu_\mu\nu_\tau} \simeq \left|\frac{y \epsilon^2}{\epsilon^2 +
2\epsilon y}
-\frac{\eta-3\epsilon}{ 1+\xi}\right|~.
\end{equation}
The second term in Eq. \eqref{eq:angle}, $(\eta-3\epsilon)/(1+\xi) \simeq
-0.53$, arises
from the charged lepton sector, while the first term, arising from the neutrino
sector is approximately equal to $\sqrt{m_{\nu_2}/m_{\nu_3}}$.  Varying
$m_{\nu_2}/m_{\nu_3}$ in the range ($1/25-1/8$), so as to be compatible with
the solar and the atmospheric neutrino oscillation data, we find, for
$y$ being positive, that
$y = (1/42.5$ to $1/44.8$), and for this range of $y$,
$\sin^22\theta_{\nu_\mu\nu_\tau}^{\rm osc} \approx (0.95-1.0)$.  (Such
a hierarchical value of $y$ goes well with the flavor symmetries
that were assumed in the Dirac mass matrices.)  We remark that the inclusion
of the first family can be carried out in the context of ESSM as in Ref.
\cite{BPW},
which would lead to eight predictions for the masses and mixings
of quarks and leptons.

\section{LEP Neutrino Counting and NuTeV Anomaly in the ESSM Framework:
Applications To Model II}

Having described the general framework, we now proceed to show how ESSM can
modify expectations for neutral current interactions at NuTeV as well
as neutrino counting at LEP. Although the analytical formulae to be derived
here in terms of the parameters $p_l$, $p'_l$, $p_\nu$ and $p'_\nu$ [see Eq.
\eqref{eq:MD}] are general, and thus apply to both Model I and Model II, we will
apply them only to Model II in this section and in the next two. The
consequences for Model I will be discussed in sec. 7.
Since the system is quite constrained by its
structure and symmetries, in particular by its description of fermion masses and
mixings, we will see that the consequences for
NuTeV and LEP, and also for charged current interactions are fully determined
within the model in terms of a single parameter $\eta_u=\kappa'_u/M_N$.
To see these, we have to go from the gauge basis, in which the mass matrices
of Eqs. \eqref{eq:MYuk} and \eqref{eq:MD} are written, to the mass eigenbasis
for the charged and
the neutral leptons.  The same transformation should then be applied to
the neutral currents and the charged currents.

The diagonalization of the mass matrices can be carried out in two steps.
Consider the Lagrangian term $\overline{\psi}_{R}^0M\psi_{L}^0$, where
$\psi_{L,R}^0$ may stand for fermions in any of the four sectors of ESSM in
the gauge basis.
The mass matrix $M$ has a form as shown in Eq. \eqref{eq:MD}.  Let us write
it as
\begin{eqnarray}
M = \left(\begin{array}{cc}0 & A \\ B & Z \end{array}\right), \nonumber
\end{eqnarray}
where $0$ is a $3 \times
3$ block matrix with all its entries equal to zero, $A$ is a $3 \times 2$
matrix, $B$ is a $2 \times 3$ matrix and $Z$ is a $2 \times 2$ matrix.
The transformation $\psi_{L,R}' = U_{L,R} \psi_{L,R}^0$, where
\begin{eqnarray}
U_{L,R} = \left(\begin{array}{cc} 1 - \frac{1}{2} \rho_{L,R}\rho_{L,R}^
\dagger &\rho_{L,R} \\ -\rho_{L,R}^\dagger & 1- \frac{1}{2}
\rho_{L,R}^\dagger
\rho_{L,R}\end{array}\right)+ O(\rho_{L,R}^3)
\end{eqnarray}
and $\rho_L^\dagger = Z^{-1} B,~\rho_R^\dagger = AZ^{-1}$ will bring
$M$ to a block-diagonal form, in which the three light chiral families get
decoupled from the two heavy ones.  The effective mass matrix from the light
sector is given as $M_{\rm light} = -AZ^{-1}B$. Such a block diagonalization
may, of course, conveniently be achieved by a two-step process involving:
(a) the transformation of the $(X,~Y,~X'~,Y')$-matrices from the gauge to the
hat-basis which decouples the electron-family from the rest [see Eq.
\eqref{eq:MD}], and (b) the subsequent block-diagonalization of the
$4\times 4$-matrix which decouples the $\mu$ and the $\tau$-families from the
vectorlike families.

For simplicity of writing, we will ignore here the mixings of the fermions in
the first (i.e., the electron) family with those of others, which could be
introduced in step (a). Ordinarily, we would expect these mixings to be
negligible if the $X$ and $X'$ matrices are sufficiently hierarchical
(i.e., $x_1\ll x_2\ll x_3$). We will discuss the possible importance of such
mixings in Appendix B.
Let us apply the procedure just described to the charged lepton sector.
The effective $\mu'$-$\tau'$ mixing matrix is found to be
\begin{eqnarray}
\label{eq:Mmutau}
M_{\mu'-\tau'} = \left[\begin{array}{cc}0 & {p_l \kappa_d \kappa_s'
/ M_E} \\
{p'_l \kappa'_d \kappa_s / M'_E} & {\kappa_d \kappa'_s / M_E}
+{\kappa'_d \kappa_s / M'_E}\end{array}\right]
\end{eqnarray}
where $M_E = M_N$ [see Eq. \eqref{eq:MYuk}] by $SU(2)_L$ symmetry. The
physical
$\mu$ and $\tau$ leptons, denoted by $\mu_{L,R}$ and $\tau_{L,R}$ are then
\begin{eqnarray}
\label{eq:L}
\mu'_{L,R} &=& c_{L,R}~\mu_{L,R} + s_{L,R}~\tau_{L,R} \nonumber \\
\tau'_{L,R} &=& -s_{L,R}~\mu_{L,R} + c_{L,R}~\tau_{L,R}
\end{eqnarray}
with $c_L = \cos\theta_L$, $s_L = \sin\theta_L$, etc.  From Eq.
\eqref{eq:Mmutau} we have
$\theta_R \simeq p_l/2$ and $\tan\theta_L \simeq p_l'/2$ (where
we have set $\kappa_s = \kappa_s', \kappa_d = \kappa_d'$).  Note that
$\mu_L-\tau_L$ mixing can be quite large in our framework [see Eq.
\eqref{eq:8newp}], while
$\theta_R$ is small (so that the correct $\mu-\tau$ mass hierarchy is
reproduced), hence the use of $\tan\theta_L$, rather than $\theta_L$,

Applying the same transformation to the relevant neutral current of the
charged lepton: ``$J_{Z^0}"=\overline{\psi}_L^0 diag(1,1,1,1,a_L)\psi_L^0$ +
$\overline{\psi}_R^0 diag (1,1,1,a_R,1)\psi_R^0$, where $\psi^0_{L,R} =
(e^0,\mu^0,\tau^0,E^0,E^{\prime 0})_{L,R}$, will lead to the following
{\it new couplings} of the $Z^0$ boson to the leptons (i.e.,
in addition to their Standard Model couplings):
\begin{eqnarray}
\label{eq:delL}
\Delta {\cal L}_{NC}^ {\rm leptons} = \frac{gZ^0}{2\cos\theta_W}(a_L-1)
\eta_d^2\left[(c_Lp'_l-s_L)^2 \overline{\mu}_L \mu_L + (s_Lp_l'+c_L)^2
\overline{\tau}_L \tau_L \right. \nonumber\\ \left. +  (c_L p_l'-s_L)
(s_Lp_l'+c_L)(\overline{\mu}_L \tau_L + \overline{\tau}_L \mu_L)\right]~.
\end{eqnarray}
Here we have defined $\eta_d = \kappa_d'/M_E$.  We shall also use related
quantities $\eta_d' = \kappa_d/M_{E'},~\eta_u = \kappa_u'/M_N$
and $\eta_u'=\kappa_u/M_{N'}$.  The new interactions of $Z^0$ with
$\mu_R$ and $\tau_R$ can be obtained from Eq. \eqref{eq:delL} by the
replacement
$L \rightarrow R$, $p_{l}'\rightarrow p_{l}$, $ \eta_d' \rightarrow \eta_d$.
Here, $a_{L,R}$ are defined as $a_{L,R} = T_3 - Q \sin^2\theta_W$.

To obtain numerical estimates of violations flavor and of universality, we
 note that to a
very good approximation, $\eta_u' = \eta_u, \eta_d'=\eta_d$, and
$\kappa_s' = \kappa_s$ (in all four sectors). We then have $\eta_d'/\eta_u'
\simeq m_b/m_t \simeq 1/60$.  Since violations of universality and
flavor-changing effects in the up and neutrino
sectors can at most be about 1-2\%, we expect
$\eta_u\lesssim 1/8$-1/10.
Such a magnitude for $\eta_u$ is quite plausible \cite{FN24}.
$\eta_d$ is then $\approx 1/500$, leading to extremely tiny effects
in the charged lepton  sectors.  For example, the ratio
$\Gamma(Z \rightarrow \mu^+\mu^-)/\Gamma(Z \rightarrow \tau^+\tau^-)$ deviates
from the Standard Model value only by about 1 part in
$10^{5}$.  The decay $Z^0 \rightarrow \mu^+\tau^-$ has a rate
proportional to $\eta_d^4 \sim 10^{-10}$.

Violations of flavor and universality, analogous to those in Eq.
\eqref{eq:delL}
 exist in the quark sector as well.  For charm and
top quarks, such effects are larger, by a factor of $(m_t/m_b)^2$ in the
amplitude, compared to the charged lepton sector.  The $Z^0 \rightarrow
c\overline{c}$ coupling deviates from the Standard Model value by an
amount given by $(p_u' \eta_u/2)^2$ or $(p_u \eta_u/2)^2$ \cite{FN26_}. Owing to
the smallness
of $|p_u|$ and $|p_u'|$ ($\approx \pm 0.1$ in the example given in the previous
section), the deviation of the rate for $Z^0 \rightarrow c \overline{c}$
from the Standard Model value is only about $10^{-4}$.

The interesting feature having its origin in the SO(10) group theoretic
structure of Eq. \eqref{eq:WrmYuk} is that {\it while non-universality in
neutral current interactions involving quarks and the
charged leptons is extremely tiny, it is not so in the neutral
lepton sector.}  This difference affects both NuTeV neutral current
cross section and LEP neutrino counting.  There are two reasons for
the difference.  First, LEP neutrino counting is sensitive to the
$Z^0 \rightarrow \nu_\tau \overline{\nu}_\tau$ coupling (while
 $Z^0 \rightarrow t\overline{t}$ is kinematically forbidden).
 Second, since the
$\nu_\mu-\nu_\tau$ oscillation angle is large, as required by
SuperKamiokande, and also as predicted by our framework, the effective
$p_l'$ parameter  is large ($\approx -1.05$), unlike the case
for charm ($p_u' \approx -0.10$).  To see the effects more concretely we
need to
diagonalize the neutral lepton mass matrix of Eq. \eqref{eq:MD}, to which
we now turn.

In addition to Eq. \eqref{eq:MD}, the three $\nu_R^i$ fields have superheavy
Majorana
masses parametrized by a matrix $M_R^\nu$.  Once the $\nu_R^i$ are
integrated out, small masses for $N_L$ and $N'_L$ fields will emerge.
(There is no direct $\nu_L^i \nu_L^j$ mass term after seesaw
diagonalization because of the structure of Eq. \eqref{eq:MD}.) Let us write
these effective mass terms (of order eV or less) as
\begin{equation}
\label{eq:newdelL}
{\cal L}_{\rm mass}^{\rm eff} = m_{11} N_L^0 N_L^0 + m_{22}N_L^{\prime 0}
N_L^{\prime 0} + 2m_{12} N_L^0 N_L^{\prime 0}~.
\end{equation}
If we denote $(M_R^\nu)^{-1}_{ij} = a_{ij}$, the mass terms are
$m_{11}=\kappa_u^2(a_{33}+2a_{23}p_\nu+a_{22}p_\nu^2)$, $m_{22} = \kappa_s^2
a_{33}$, $m_{12} = \kappa_u\kappa_s(a_{33}+p_\nu a_{23})$.  The $N_L^0$ and
$N_L^{\prime 0}$ fields have Dirac masses (by combining with $N_R^0$ and
$N^{\prime 0}_R$ respectively) of order few hundred GeV; they
also possess non-diagonal
Dirac mass mixing terms involving the light neutrinos.  Upon identifying
the light components, Eq. \eqref{eq:newdelL} will generate small Majorana
masses of the standard left-handed neutrinos.

We can block diagonalize the Dirac mass matrix of the neutral leptons which
is obtained from Eq. \eqref{eq:MD} after integrating out the superheavy
$\nu_R^i$ fields.
This can be done by applying a unitary transformation on $(\nu_\mu^0,
\nu_\tau^0,
N_L^0, N_L^{'0})$ fields.  Note that the $(N_R,~N_R'$) fields do not mix
 with the
light neutrinos, since $\nu_R^i$ are superheavy. [We remind the reader that
(for the sake of simplicity) we ignore for the present the mixing of the
fermions of the first family with those of the vectorlike families. Such
mixings will have a very small effect on LEP neutrino-counting, which we will
incorporate, and likewise on NuTeV measurements. Possible consequences of such
mixing are discussed in Appendix B.]
Define
\begin{eqnarray}
\label{eq:some}
a&=&p_\nu' \eta_u,~ b= \eta_u, ~c=\kappa_s'/M_N = \eta_s,
\nonumber \\
N_2 &=& \sqrt{1+a^2+b^2},~N_3=\sqrt{1+b^2+c^2},~N_4 =
\sqrt{(1+a^2)(1+b^2)+c^2}~.
\end{eqnarray}
The transformation $(\nu_2', \nu_3', \nu_4', \nu_5')^T = U^\nu (\nu_\mu^0,
\nu_\tau^0, N, N')^T_L$, where
\begin{eqnarray}
\label{eq:Unu}
U^\nu = \left(\begin{array}{cccc}\frac{N_3}{N_4} & -\frac{ab}{N_3 N_4} &
 \frac{abc}{N_3 N_4} & -\frac{a(1+c^2)}{N_3 N_4} \\
0 & \frac{1}{ N_3} & -\frac{c}{N_3} & -\frac{b}{N_3} \\
-\frac{abc}{ N_2 N_4} & \frac{c(1+a^2)}{N_2 N_4} & \frac{N_2}{ N_4} &
-\frac{bc}{ N_2 N_4} \\
\frac{a}{ N_2} & \frac{b }{ N_2} & 0 & \frac{1}{ N_2}\end{array}\right)
\end{eqnarray}
block diagonalizes the Dirac mass entries, so that there is no mixing
between the massless states ($\nu_2', \nu_3'$) and the massive states
($\nu_4', \nu_5'$).  From Eq. \eqref{eq:Unu}, one can read off the light
mass eigenstate
components in the original fields defined in the gauge basis.  For example,
$\nu_\mu^0 = (N_3/N_4) \nu_2' + ...$, $N_L = [abc/(N_3N_4)] \nu_2' - (c/N_3)
 \nu_3'
+ ...$, etc, where the dots denote the heavy components, which we drop since
they are not kinematically accessible to NuTeV and LEP. Once these heavy
components are dropped, the resulting states are not normalized to unity
and it is this feature that is relevant to the NuTeV anomaly and LEP
neutrino counting.

As a digression, we may mention that when the light neutrino mass matrix
resulting from Eq. \eqref{eq:newdelL} is written in terms of
$(\nu_2',\nu_3')$ fields,
it is given approximately (by setting $a_{33} = 0$, and $a_{23} \ll a_{22}$,
both of which follow from the form of $M_R^\nu$ given before) by,
\begin{eqnarray}
\label{eq:Mnul}
M_\nu^{\rm light} \simeq \left(\begin{array}{cc} 0 & a_{23} p_\nu \cr
a_{23} p_\nu & a_{22} p_\nu^2 +4a_{23} p_\nu\end {array}\right)
\frac{\kappa_u^2 \kappa_s^2}{ M_N^2}~.
\end{eqnarray}
Eq. \eqref{eq:Mnul} is of course completely equivalent to Eq. \eqref{eq:Mnu},
except for having a reparametrization, and thus preserves the prediction of
large $\nu_\mu$-$\nu_\tau$ oscillation angle [see discussion below Eq.
\eqref{eq:Mnu}].

Having identified the light neutrino states through Eq. \eqref{eq:Unu}, we
can calculate
the correction to neutrino counting at LEP.  In the gauge basis, the $Z^0$
coupling is given by $[g/(2 \cos\theta_W)]Z^0[\overline{\nu}_e^0 \nu_e^0
+ \overline{\nu}_\mu^0 \nu_\mu^0 + \overline{\nu}_\tau^0 \nu_\tau^0 +
\overline{N}_L^0 N_L^0 + \overline{N}_R^0 N_R^0]$.  The last term does not
affect
$N_\nu$ (the number of light neutrinos counted at LEP), since $N_R$ is
heavier than $Z$, and since it has no mixing with the light neutrinos.
Applying the transformation of Eq. \eqref{eq:Unu} we find that owing to mixing
of the current eigenstates $\nu_\mu$ and $\nu_\tau$ with $N'$, the couplings
of $Z^0\rightarrow \bar \nu_\mu\nu_\mu$ and $Z^0\rightarrow \bar \nu_\tau
\nu_\tau$ are reduced compared to their standard model values by the factors
$(1-\delta_\mu)$ and $(1-\delta_\tau)$ respectively, where
\begin{eqnarray}
\label{eq:deltas}
\delta_\mu\equiv\eta^2_u(c_Lp'_\nu-s_L)^2;\quad
\delta_\tau\equiv\eta^2_u(c_L+s_Lp'_\nu)^2.
\end{eqnarray}
In this description, we have ignored corrections of order
$\eta_u^4\lesssim 10^{-4}$ [see Eqs. \eqref{eq:Lcc}-\eqref{eq:LZcc} for an
accurate description]. The corresponding angles for mixing of $\nu_\mu$ with
$\nu_\tau$ with $N'$ are given by $\theta_{\nu_\mu N'}\approx\sqrt{\delta_\mu}$
and $\theta_{\nu_\tau N'}\approx\sqrt{\delta_\tau}$. Now, in general, $\nu_e$
also mixes with $N'$ owing to (a) the transformation of the $X'$ matrix to the
hat-basis discussed in sec. 2 and in Appendix B, which introduces
$\nu_e$-$\nu_\mu$ mixing given by $\theta'_{\nu_e \nu_\mu}$
(see Eq. \eqref{eq:B7}), and (b) $\nu_\mu$-$N'$ mixing as mentioned above. Thus
 we get:
$\theta_{\nu_e N'}\approx (\theta'_{\nu_e \nu_\mu})(\theta_{\nu_\mu N'})
\approx  (\theta'_{\nu_e \nu_\mu})\sqrt{\delta_\mu}$. This mixing would reduce
the $Z^0\rightarrow \bar \nu_e\nu_e$ coupling by the factor $(1-\delta_e)$,
where $\delta_e\equiv (\theta'_{\nu_e \nu_\mu})^2\delta_\mu\lesssim
\delta_\mu/16$, if $\theta'_{\nu_e \nu_\mu}$ is set to be $\lesssim 1/4$
(see Appendix B). Thus ordinarily $\delta_e$ is expected to be extremely small.
we will still exhibit it in the equations for the sake of generality. The
{\it net} neutrino counting observed at LEP involving $Z^0$-decays into
$\bar\nu_i\nu_i$-pairs is then given by:
\begin{equation}
\label{eq:NLEP}
N_\nu({\rm LEP}) = 3 - 2(\delta_\mu+\delta_\tau+\delta_e)=
3-2\eta_u^2(1+p^{\prime 2}_\nu)-2\delta_e,
\end{equation}
This expression for $N_\nu$ holds for both Models I and II. For numerical
purposes, we will apply it in this section only to Model II.
The experimental value from LEP is $N_\nu = 2.9841 \pm 0.0083$ \cite{27}.
We see that ESSM leads to a {\it reduction} in $N_\nu$, which is in good
 agreement with the LEP data. Setting $p_\nu' = 0.3$ for Model II
[see Eq. (8)], and $\eta_u = 1/10$-1/15, we have $N_\nu = (2.9782$
to $2.9903)-2\delta_e$, where the $(2\delta_e)$-term makes a negligible
contribution: ($2\delta_e\lesssim 0.0005$ for $\eta_u$ as stated above).
The suggested two sigma deviation in $N_\nu$ measured at LEP compared to
Standard Model may thus be taken as a hint for $\nu_\tau$-$N'$ and
$\nu_\mu$-$N'$ mixings.  The LEP value for $N_\nu$ would imply a magnitude for
$\eta_u \approx (1/10-1/15)$, which can
then be used to predict deviations in the other experiments, such as NuTeV.
Theoretically, values of $\eta_u\approx 1/5$-1/20 are perfectly plausible
(see Ref. \cite{FN24,FN26_})

There are modifications in the charged current interactions as well, which
is straightforward to compute:
\begin{equation}
\label{eq:Lcc}
{\cal L}_{cc}=\frac{g}{ \sqrt{2}} W^+[ (1-\delta_e/2)\overline{\nu}_e e_L +
(1-\delta_\mu/2) \overline{\nu}_\mu \mu_L +
(1-\delta_\tau/2)\overline{\nu}_\tau \tau_L] + h.c.
\end{equation}
Here we {\it define} $\nu_\mu$ as the normalized state that couples to
$\mu_L^-W^+$ and similarly $\nu_\tau$ as the normalized state that couples
to $\tau_L^-W^+$.  In terms of $\nu_2'$ and $\nu_3'$, $\nu_\mu$ is
given as
\begin{equation}
\label{eq:20}
\nu_\mu = \cos\phi~\nu_2' + \sin\phi~\nu_3'
\end{equation}
where $\sin\phi = -s_L(1-b^2/2)-(s_L/2)(b s_L - a c_L)^2$.  The state
orthogonal to $\nu_\mu$, viz., $\hat{\nu}_\tau = -\sin\phi ~\nu_2'+
\cos\phi~\nu_3'$, is not exactly $\nu_\tau$.  (Thus, $\nu_\mu$
produced in $\pi$ decays can produce $\tau$ leptons, but numerically
this cross section is very small, being proportional to $
\eta_u^4 \sim 10^{-5}$.)

In order to see how the model accounts for the NuTeV neutral current
anomaly, it is useful to rewrite the $Z^0 \overline{\nu}_i'\nu_j'$ interaction
in terms of the current eigenstate $\nu_\mu$ and the state orthogonal to it
($\hat{\nu}_\tau$).  Including the first family, it is
given as
\begin{eqnarray}
\label{eq:LZcc}
{\cal L}_{NC}^Z &=& \frac{g}{2 \cos\theta_W}Z^0[\bar{\nu}_\mu \nu_\mu
(1-\delta_\mu) + (\bar{\nu}_\mu \hat{\nu}_\tau + \bar{\hat{\nu}}_\tau \nu_\mu)
\{(b^2-a^2)c_Ls_L-(c_L^2-s_L^2)ab\} \nonumber \\
&+&\bar{\hat{\nu}}_\tau \hat{\nu}_\tau(1-\delta_\tau)
+\bar\nu_e\nu_e(1-\delta_e)]~.
\end{eqnarray}

Using Eq. \eqref{eq:Lcc}, and including radiative corrections, the Fermi
coupling $G_\mu$ is given in our model (including radiative corrections) by
\begin{equation}
\label{eq:Gmu}
\frac{G_\mu}{ \sqrt{2}}^{\rm ESSM}=\frac{g^2}{8 m_W^2}(1+
\Delta r_W^{(\nu_\mu,l)})(1-\delta/2),
\end{equation}
where $\delta\equiv\delta_\mu+\delta_e\approx\delta_\mu$ and
$\Delta r_W^{(\nu_\mu,l)}$ denotes
electroweak radiative corrections to $\mu$-decay, which is usually denoted by
$\Delta r$ in the literature \cite{degrassi,marciano2,erler}. Note that
by definition, it is the right side of Eq. \eqref{eq:Gmu} which is determined
by the observed muon decay rate: thus $G_\mu^{\rm ESSM}=G_\mu^{\rm obs}$. The
amplitude for the CC process $(\nu_\mu N\rightarrow \mu X)$ is then given by
$(G_\mu^{\rm obs}/\sqrt{2})(1+\Delta r'_W)$ in both the SM and ESSM. Here,
$\Delta r'_W$ denotes the difference \cite{FN37} between the radiative
correction $(\Delta r_W^{(\nu_\mu,h)})$ to the amplitude for $(\nu_\mu N
\rightarrow \mu X)$ and that for $\mu$-decay; thus $\Delta r'_W=
\Delta r_W^{(\nu_\mu,h)}-\Delta r_W^{(\nu_\mu,l)}$. Since $\Delta r'_W$ is the
same for both ESSM and the SM, the CC cross sections are the same in the two
cases: $[\sigma_{CC}^{\nu(\bar\nu)}]_{\rm ESSM}=
[\sigma_{CC}^{\nu(\bar\nu)}]_{\rm SM}$.

On the other hand, the NC cross
section $\sigma(\nu_\mu N \rightarrow \nu X)$ will be modified. To see this,
we use the fact that in ESSM, the amplitude for the process
$\nu_\mu q_i\rightarrow \nu_\mu q_i$ (including radiative corrections) is
proportional to:
\begin{eqnarray}
\label{eq:Aqiqi}
A(\nu_\mu q_i\rightarrow \nu_\mu q_i)^{\rm ESSM}&\propto&
\frac{g^2}{8m^2_Z\cos^2\theta_W}(1+\Delta r^i_Z)a_i(1-\delta_\mu)\nonumber\\&=&
(G_\mu^{\rm obs}/\sqrt{2})a_i(1+\Delta r^i_Z-\Delta r_W^{(\nu_\mu,l)})
(1-\delta'/2),
\end{eqnarray}
where $\delta'\equiv\delta_\mu-\delta_e\approx\delta_\mu$.
Here $a_i=(I_{3L}-Q\sin^2\theta_W)_i$ and $\Delta r^i_Z$ denotes the radiative
correction to the amplitude for $\nu_\mu q_i\rightarrow \nu_\mu q_i$. In
getting the second step of Eq. \eqref{eq:Aqiqi}, we have used Eq.
\eqref{eq:Gmu} and the on-shell relation $m^2_W/(m^2_Z\cos^2\theta_W)=1$, where
$m_W$ and $m_Z$ denote the physical masses of $W$ and $Z$, respectively.
Since the radiative corrections to the NC processes, discussed in detail in
Ref. \cite{Marciano}, are the same for the SM and ESSM (barring small
differences owing to differences between predictions for the Higgs mass in the
two cases \cite{FN39}), we obtain (using Eq. \eqref{eq:Aqiqi}):
\begin{eqnarray}
\label{eq:ss}
\sigma(\nu_\mu N\rightarrow \nu X)^{\rm ESSM}=
\sigma(\nu_\mu N\rightarrow \nu X)^{\rm SM}(1-\delta_\mu),
\end{eqnarray}
where we have set $\delta'=\delta_\mu$
Notice that the neutral current cross section is {\it reduced} compared to its
SM value. Using the values corresponding to Model II -- that is $p'_\nu=0.3$,
$s_L/c_L=-0.526$ [fixed by our considerations
of fermion masses and mixings, see Eqs. \eqref{eq:4matrices}-\eqref{eq:8p}],
and $\eta_u=1/11.6$-1/13.3, we get $\delta_\mu=0.003$ to 0.004. While this range
for $\eta_u$ and equivalently for $\delta_\mu$ is quite reasonable \cite{FN24},
 we stress that we can not choose $\delta_\mu$ to be much higher owing to
constraints
from charged-current universality, discussed below. We thus obtain a deviation:
$\sigma(\nu_\mu N\rightarrow \nu X)^{\rm ESSM}/
\sigma(\nu_\mu N\rightarrow \nu X)^{\rm SM}\approx 1-$(0.3 to 0.4)\%, in
Model II.
Considering that NuTeV measurements may be interpreted as a reduction of the
NC cross section by about $ (1.2\pm 0.4\%)$, compared to the SM
\cite{Farland}), keeping the CC cross section and $\sin^2\theta_W$ unchanged,
we see that ESSM would reduce the deficit in the NC cross section to
about $0.8\pm 0.4\%$. This can
partially account for the NuTeV anomaly by reducing it from a $3\sigma$ to a
$2\sigma$-effect. This is of course statistically quite significant. It
remains to be seen whether a portion or even a bulk of the NuTeV anomaly can
be attributed to a Standard Model-based QCD and other
effects, considered by several authors, including the experimenters
\cite{Strumia}.

We stress the intimate quantitative link between the reduction in LEP
neutrino-counting $N_\nu$ and that in the NuTeV cross section [see Eqs.
\eqref{eq:NLEP} and \eqref{eq:ss}], which emerges because all the relevant
parameters are fixed owing to our considerations of fermion masses and
mixings.  It is worth noting that owing to hierarchical masses of the
three families, and thus nonuniversal mixings of ($\nu_e,~\nu_\mu$ and
$\nu_\tau$) with $N'$, the reduction in $N_\nu$ from 3 is {\it not} simply
three times the relative reduction in $\sigma(\nu_\mu N \rightarrow \nu X)$
[compare Eqs. \eqref{eq:NLEP} and \eqref{eq:ss}].

\section{Changes in $m_H$ and $m_W$ in Model II due to modification in $G_\mu$}

It is worth noting that the tree--level modification of the expression
for $G_\mu$ containing the $\eta_u^2$-term [see Eq. (23)] can have a
testable consequence as follows.  The familiar Standard Model expression,
in the on-shell scheme, for $G_\mu$, including electroweak radiative
corrections \cite{Marciano,degrassi,marciano2,erler}, is now modified to
\begin{equation}
\label{eq:Gmuu}
G_\mu = \frac{\pi \alpha}{\sqrt{2}\; m_W^2(1-m_W^2/m_Z^2)
(1-\Delta r +\delta/2)}~,
\end{equation}
where $\delta/2=(\delta_\mu+\delta_e)\approx\delta_\mu/2=
\eta_u^2(c_L p_\nu'-s_L)^2/2 \approx 0.0015-0.0020$ (for Model II, see above),
and we have used $(\sin^2\theta_W)_{\rm on~shell}\equiv
1-m_W^2/m_Z^2$ \cite{Marciano,newrad}.
Now, $\Delta r$ depends on $m_t$ and $m_H$.  Using the CDF/D0 value
of $m_t = 174.3 \pm 5.1$ GeV, and $\alpha^{-1}(m_Z) = 128.933(21)$
\cite{davier},
one obtains \cite{marciano2} $\Delta r = (0.03402,~0.03497,~0.03575~0.03646)$
for $m_H = (75,~100,~125$ and $150$) GeV with an uncertainty $\delta (\Delta r)
= \pm 0.0020 \pm 0.0002$, where the first uncertainty is primarily due to that
in $m_t$ by $\pm 5.1$ GeV.
Note that $\Delta r$ increases with $m_H$, for a fixed $m_t$.  Defining
$(\Delta r)_{\rm eff} \equiv \Delta r - \delta/2$, we see that for
any given reference point of $(m_t,~m_H)$, and thus for $\Delta r$,
$(\Delta r)_{\rm eff}$ is necessarily lower than $\Delta r$.  Eq.
\eqref{eq:Gmuu} then
tells us that, with a lower $(\Delta r)_{\rm eff}$, compared to $\Delta r$,
{\it ESSM will predict a higher value of $m_W$}, compared to that of the SM,
so that $G_\mu$ and $m_Z$ are held fixed, which are measured most accurately.
To be specific, for any given choice of ($m_t$ and $m_H$), in order that
$G_\mu$ and $m_Z$ may be held fixed, Eq. \eqref{eq:Gmuu} imposes the condition
\begin{eqnarray}
\label{eq:mm}
\hat m_W^2(1-\hat m_W^2/m_Z^2)(1-\Delta r)=
m_W^2(1-m_W^2/m_Z^2)(1-\Delta r_{\rm eff}),
\end{eqnarray}
where $\hat m_W$ and $m_W$ denote the mass of the $W$-boson for the SM and
ESSM, respectively. Defining $m_W=(m_W)_{\rm ESSM}\equiv \hat m_W+\delta m_W$,
it is easy to verify that Eq. \eqref{eq:mm} yields (to a very high accuracy):
\begin{eqnarray}
\label{eq:dm}
(\delta m_W)_{{\rm fixed~}(m_t,m_H)}=\left(\frac{m_Z^2-m_W^2}{2m^2_W}\right)
\left[1-\frac{m_Z^2-m_W^2}{m^2_W}\right]^{-1}m_W(\delta/2)\approx (0.20)m_W
(\delta/2).
\end{eqnarray}
Thus, the shift $\delta m_W$ (for fixed $m_t$ and $m_H$) is positive. Taking
$(\delta/2)=0.0015$-0.0020 (for Model II), we obtain:
$m_W=(m_W)_{\rm SM}+$(24-32) MeV, for any fixed value of ($m_t$ and $m_H$)
\cite{FN44}.

Alternatively, noting that $m_H$ is fixed by $\Delta r$ in both the SM and
the ESSM for a given $m_t$, if one wishes to fix $m_W$ [say within about one
sigma of its
measured value $m_W^{\rm exp} = 80.446 \pm 0.040$ GeV (LEP data)], which
in turn fixes $(\Delta r)_{\rm eff}$ via Eq. \eqref{eq:Gmuu},
{\it  one would be led to
predict (using $\Delta r = (\Delta r)_{\rm eff} + \delta/2$)
a higher value of $m_H$ in ESSM compared to that in the SM for any
fixed value of $m_t$.}

To be concrete,
using $m_t = 174.3$ GeV, and choosing $m_H = 125$ GeV (in accord with
LEP search that shows $m_H \ge 115$ GeV \cite{lephiggs}), one gets
$(m_W)_{SM} = 80.372$ GeV \cite{marciano2}, which is about 1.8 sigma
below the measured value, whereas one obtains $(m_W)_{ESSM} = 80.396-
80.404$ GeV, for $\delta/2 = 0.0015-0.002$ (in Model II), which is in better
agreement with the observed value.  Alternatively, with $m_t = 174.3$ GeV,
if one chooses to fix $m_W = 80.401$ GeV (1.1 sigma below the measured value),
one would predict (as a central value) $(m_H)_{SM} \approx 75$ GeV
\cite{marciano2,langacker2},
whereas $(m_H)_{ESSM}\approx 125$ GeV (for $\delta/2=0.0017$), in Model II.

In general, for any fixed $m_t$, the replacement of $\Delta r$ by
$\Delta r_{\rm eff}=\Delta r-\delta/2$ would, of course, correspond to changes
in both $m_H$ and $m_W$. A few sample cases, exibiting the SM and ESSM
predictions for ($m_H$ and $m_W$), based on Eq. \eqref{eq:Gmuu} are exibited
in Table I, for the case of Model II.
($m_t=174.3$ GeV is held fixed \cite{FN46}).\\

\noindent\begin{tabular}{c||c|c||c|c||c|c||c|c}
case&\multicolumn{2}{c||}{(a)}&\multicolumn{2}{c||}{(b)}&
\multicolumn{2}{c||}{(c)}
&\multicolumn{2}{c}{(d)}
\\&\multicolumn{2}{c||}{Fix $m_W$}&\multicolumn{2}{c||}{Fix $m_H$}&
\multicolumn{2}{c||}{Vary ($m_H$, $m_W$)}
&\multicolumn{2}{c}{Vary ($m_H$, $m_W$)}\\
\hhline{~|-|-|-|-|-|-|-|-}
&$m_H$&$m_W$&$m_H$&$m_W$&$m_H$&$m_W$&$m_H$&$m_W$\\
\hline
SM&75&80.401&125&80.372&115&80.377&100&80.385\\
\hline
ESSM&116.5-135&80.401&125&80.396-80.404&123-140&80.397&108-125&80.405
\end{tabular}\\

{\bf Table I.} Predictions for $m_H$ and $m_W$, based on Eq. \eqref{eq:Gmuu},
with
$m_t=174.3$ GeV and $\delta/2=$0.0015-0.0020. The masses are in units of GeV.
The ESSM-predictions given above are for Model II.\\

We thus see that, ESSM (for Model II) would generically predict higher values
of $m_H$ and/or
of $m_W$ compared to that in the SM \cite{FN47}. While this feature seems to
be in better
accord with observations at present, its validity can be ascertained  once
(a) $m_t$ is measured to within $1-2$ GeV, (b)
$m_W$ is measured to better than $10-20$ MeV accuracy, and (c) if the light
Higgs particle (say with a mass $\lesssim 150$ GeV)
is discovered in future searches at the Tevatron, the LHC, and
the NLC.

\section{Charged Current Processes In Model II}

We now turn to the question of universality in leptonic charged current
processes.
The flavor dependence of the charged current couplings predicted by our
framework, Eq. \eqref{eq:Lcc} , will lead to nonuniversality in leptonic
 decays,
correlated with the $\nu_\mu$-nucleon neutral current cross section
measured at NuTeV, as well as neutrino counting at LEP.

The leptonic decays of $\pi^+$ mesons provide a sensitive probe of $e$-$\mu$
universality.  In the Standard Model, the branching ratio $R_{e/\mu}^{\rm SM}
\equiv
\Gamma(\pi^+ \rightarrow e^+ \nu_e)/\Gamma(\pi^+ \rightarrow \mu^+ \nu_\mu)$
has been computed quite accurately, including radiative corrections to be
\cite{marcianosirlin}
$R_{e/\mu}^{\rm SM} = (1.2352 \pm 0.0004) \times 10^{-4}$.  In our framework,
this prediction is modified to [see Eq. \eqref{eq:Lcc}]
\begin{equation}
R_{e/\mu}^{\rm ESSM}=R_{e/\mu}^{\rm SM}(1+\delta'),
\end{equation}
where $\delta'=\delta_\mu-\delta_e\approx\delta_\mu$
The PSI experiment
\cite{psi} measures this ratio to be $R_{e/\mu}^{\rm exp-PSI} = (1.2346 \pm
0.0050)\times 10^{-4}$, whereas the TRIUMF experiment
\cite{triumf} finds it to be
$R_{e/\mu}^{\rm exp-TRIUMF} = (1.2285 \pm 0.0056) \times 10^{-4}$.
If we choose $\delta_\mu=\eta_u^2(c_L p_\nu'-s_L)^2 =$0.3 to 0.4\% (for Model
II), so that the deviation
from the Standard Model in $\nu_\mu$--nucleon neutral current cross section at
NuTeV is 0.3 to 0.4\%, we have $R_{e/\mu}^{\rm ESSM} = (1.2389$ to
$1.2401) \times 10^{-4}$.  This
value is about 0.86 to 1.1 sigma above the PSI measurement, and about
1.8 to 2.1 sigma
above the TRIUMF measurement.  We consider these deviations arising within
Model II, although not
insignificant for the TRIUMF experiment, to be within acceptable range. Modest
improvements in these measurements can
either confirm or entirely exclude our explanation of the
indicated NuTeV anomaly.

It should also be mentioned that $e-\mu$ universality is well tested
in $\tau^+ \rightarrow e^+ \nu_e \bar{\nu}_\tau$ versus
$\tau^+ \rightarrow \mu^+ \nu_\mu \bar{\nu}_\tau$ decays as well.
The effective Fermi coupling strength $G_{\tau e} $ and $G_{\tau \mu}$
characterizing these decays are in the ratio \cite{GteGtu}
 $G_{\tau e}/G_{\tau \mu}
= 0.9989 \pm 0.0028$.  Our framework predicts it to be
$(G_{\tau e}/G_{\tau \mu})^{\rm ESSM} = (G_{\tau e}/G_{\tau \mu})^{\rm SM}[1+
\delta_\mu/2]$.  Using the correction factor in this
ratio to be 0.15 to 0.2\%
(so that deviation at NuTeV is 0.3 to 0.4\%), we find the deviation from
experiment to be
at the level of 0.9 to 1.1 sigma, which is quite acceptable. In Appendix A, we
discuss the modifications in ESSM for $V_{ud}$ and $V_{us}$ and the resulting
implications for the unitarity of the CKM matrix. The results in this regard
for Model I and Model II are compared in the next section.

\section{Results for Model I: Comparison with Model II}

As noted in the introduction, pattern II for fermion masses and mixings,
presented in sec. 3, is an interesting variant of pattern I \cite{BP(g-2),BPW}.
Both of these arise within the SO(10)-framework and possess the same degree
of success as regards their predictions for fermion masses and mixings. They
can, however, be clearly distinguished from each other if they are extended to
ESSM. The extensions of patterns I and II to ESSM are referred to as Models I
and II,
respectively. So far, in secs. 4, 5 and 6, we have considered the
consequences of Model II only. We now turn attention to Model I.

Following Ref. \cite{BPW} we recall that pattern I for the fermion masses for
the $\mu$-$\tau$ sector is given by:
\begin{eqnarray}
\label{eq:4M}
\begin{array}{ccc}& \begin{array}{cc} c_L & t_L \end{array}&\\
M_u=&\left(\begin{array}{cc} 0 & \sigma-\epsilon\\ \sigma+\epsilon & 1
\end{array}\right)&{\cal M}_U^0;
\end{array}\quad
\begin{array}{ccc}& \begin{array}{cc} s_L & b_L \end{array}&\\
M_d=&\left(\begin{array}{cc} 0 & \eta-\epsilon\\ \eta+\epsilon & 1
\end{array}\right)&{\cal M}_D^0;
\end{array}\nonumber\\
\begin{array}{ccc}& \begin{array}{cc} \nu_L^\mu & \nu_L^\tau \end{array}&\\
M_\nu^D=&\left(\begin{array}{cc} 0 & \sigma+3\epsilon\\ \sigma-3\epsilon & 1
\end{array}\right)&{\cal M}_U^0;
\end{array}\quad
\begin{array}{ccc}& \begin{array}{cc} \mu_L & \tau_L \end{array}&\\
M_l=&\left(\begin{array}{cc} 0 & \eta+3\epsilon\\ \eta-3\epsilon & 1
\end{array}\right)&{\cal M}_D^0.
\end{array}
\end{eqnarray}
Compare these with pattern II given in Eq. \eqref{eq:4matrices}. These, together
with their extension to include the first family, yield seven predictions, all
in good accord with observations \cite{BPW}. The parameters $\eta$, $\epsilon$,
$\sigma$, ${\cal M}_U^0$ and ${\cal M}_D^0$ are determined to be:
\begin{eqnarray}
\label{eq:param}
\left.\begin{array}{ccc}
\eta\approx -0.151,&\epsilon\approx 0.095,&\sigma\approx -0.110,\\
{\cal M}_U^0\approx 110~\mbox{GeV},& {\cal M}_D^0\approx 1.5~\mbox{GeV}.&
\end{array}\right\}\mbox{ (Model I)}
\end{eqnarray}
Embedding of pattern I in ESSM, which we call Model I, is discussed in Ref.
\cite{BP(g-2)}. Using Eq. \eqref{eq:param}, the corresponding $p_f$ and $p'_f$
parameters defined in sec. 2 (see e.g., Eq. \eqref{eq:MD}) are given by:
\begin{eqnarray}
\left.\begin{array}{cc}
p_u=2(\sigma-\epsilon)=-0.41;& p'_u=2(\sigma+\epsilon)=-0.03;\\
p_d=2(\eta-\epsilon)=-0.492;& p'_d=2(\eta+\epsilon)=-0.112;\\
p_\nu=2(\sigma+3\epsilon)=0.350;& p'_\nu=2(\sigma-3\epsilon)=-0.79;\\
p_l=2(\eta+3\epsilon)=0.268;& p'_l=2(\eta-3\epsilon)=-0.872.
\end{array}\right\}\mbox{ (Model I)}
\end{eqnarray}
Compare these with the corresponding values for Model II given in Eqs.
\eqref{eq:8p} and \eqref{eq:8newp}. Note the crucial difference in magnitude
{\it and} sign between the values of the parameter $p'_\nu$ in the two models:
$p'_\nu=-0.79$ (for Model I), while $p'_\nu=+0.30$ (for Model II). The values
of the parameter $p'_l$ (which determines the mixing angle $\tan\theta_L\approx
p'_l/2$ (see Eq. \eqref{eq:L}) on the other hand are quite similar in the two
models: $p'_l=-0.87$ (for Model I), while $p'_l=-1.05$ (for Model II). These
two parameters together (whose magnitudes and relative signs are completely
fixed within each model) lead to a host of differences between the predictions
of the two models.

For comparison purposes, the values of certain relevant quantities are noted
below for the Model I and II [we recall the definitions $\delta_\mu=\eta_u^2
(c_Lp'_\nu-s_L)^2$, $\delta_\tau=\eta_u^2(c_L+s_Lp'_\nu)^2$ given in Eq.
\eqref{eq:deltas}, and the relations $\tan\theta_L\approx p'_l/2$
(see Eq. \eqref{eq:L}) and
$N_\nu=3-2(\delta_\mu+\delta_\tau)=3-2\eta_u^2(1+p^{\prime 2}_\nu)$
(see Eq. \eqref{eq:NLEP}), where we have neglected $\delta_e$]:
\begin{eqnarray}
\begin{array}{ccc}
\mbox{\underline{Model I}}&\quad\quad\quad&\mbox{\underline{Model II}}\\
p'_l=-0.87,\quad p'_\nu=-0.79 && p'_l=-1.05,\quad p'_\nu=0.30\\
\theta_L\approx \tan^{-1}(p'_l/2)\approx 156.49^0&& \theta_L\approx 152.3^0\\
\delta_\mu\approx\eta_u^2(0.1060)&&\delta_\mu\approx\eta_u^2(0.5335)\\
\delta_\tau\approx\eta_u^2(1.5182)&&\delta_\mu\approx\eta_u^2(0.5564)\\
N_\nu=3-\eta_u^2(3.248)&&N_\nu=3-\eta_u^2(2.18)
\end{array}
\end{eqnarray}
The following features are worthnoting:

(1) For any choice of $\eta_u$,
$\delta_\mu$ is suppressed in Model I relative to that in Model II by about a
factor of 5. This is the main reason for the differences between the predictions
of the two models (as we elaborate below).

(2) $\delta_\tau$ on the other hand
is enhanced in Model I relative to Model II by about a factor of 2.7. This
would lead to differences between the two models in their predictions for the
tau lifetime.

(3) Despite the differences between the two models as regards
the values of $\delta_\mu$ and $\delta_\tau$, in particular despite the
smallness of $\delta_\mu$ in Model I, it is rather interesting that both
models lead to a substantial decrease in $N_\nu$ from 3 (see elaborations
below), in good accord with the LEP data \cite{27}. This is because the decrease
in $N_\nu$ depends on the sum $(\delta_\mu+\delta_\tau)$ and the smallness of
$\delta_\mu$ in Model I is fully compensated by the largeness of $\delta_\tau$.
Note that these features are fixed even qualitatively by the nature of the
SO(10)-based patterns [Eqs. \eqref{eq:4matrices} and \eqref{eq:4M}] and our
consideration of the masses and mixings of quarks and leptons.

The results for Model II pertaining to $N_\nu$, the NuTeV-measurements and other
entities have been discussed in detail in secs. 4, 5 and 6. We now present the
corresponding results for Model I in table II, where the results for Model II
are also listed for the sake of a comparison. Following our discussion in
Appendix A, we have included, in table II, the predictions for the effective
values of $|V_{ud}|$ and $|V_{us}|_{\rm unitarity}$ in ESSM.\\

\noindent
\begin{table}
{\bf Table II.} Results for Models I and II of ESSM. The deviations from SM
predictions for the NuTeV measurements involving the ratio of NC to CC cross
sections are simply given by deviations of the ratio
$(\sigma_{\rm NC}^{\rm ESSM}/\sigma_{\rm NC}^{\rm SM})$ from unity, because
$\sigma_{\rm CC}^{\rm ESSM}=\sigma_{\rm CC}^{\rm SM}$, see discussions
following Eq. \eqref{eq:Gmu}. The quantity
$\Delta us$ in the last row stands for 0.0021, see Eq. (40).\\

\rotatebox{90}{
\begin{tabular}{c||c|c||c|c||c|c||c|c}
$\eta_u$&\multicolumn{2}{c||}{1/12}&\multicolumn{2}{c||}{1/15}&
\multicolumn{2}{c||}{1/18}&\multicolumn{2}{c}{1/20}\\
\hline
Models&I&II&I&II&I&II&I&II\\
\hline
$\delta_\mu$&0.00074&0.00370&0.00047&0.00237&0.00033&0.00165&0.00027&0.00133\\
\hline
$\delta_\tau$&0.0105&0.00386&0.00675&0.00247&0.00469&0.00171&0.00380&0.00139\\
\hline
$N_\nu=3-2(\delta_\mu+\delta_\tau)$&2.9775&2.9850&2.9855&2.9903&2.9900&2.9933&
2.9919&2.9946\\
\hline
$(\sigma_{\rm NC}^{\rm ESSM}/\sigma_{\rm NC}^{\rm SM})-1=-\delta_\mu$
&-0.074\%&-0.37\%&-0.05\%&-0.24\%&-0.03\%&-0.17\%&-0.027\%&-0.13\%\\
\hline
$(R_{e/\mu}^{\rm ESSM}/R_{e/\mu}^{\rm SM})-1=+\delta_\mu$
&0.074\%&0.37\%&0.05\%&0.24\%&0.03\%&0.17\%&0.027\%&0.13\%\\
\hline
$(\tau_\tau^{\rm ESSM}/\tau_\tau^{\rm SM})-1\approx\delta_\tau+\delta_\mu/5.7$
&1.06\%&0.45\%&0.68\%&0.29\%&0.47\%&0.20\%&0.38\%&0.16\%\\
\hline
$|V_{ud}^{\rm ESSM}/(V_{ud})_{\rm SM}^{\rm expt}|-1\approx -\delta_\mu/2$
&0.00037&0.00185&0.00024&0.00119&0.00016&0.00082&0.000135&0.00066\\
\hline
$|V_{us}^{\rm ESSM}|_{\rm unitarity}$&0.22850
&0.23470&0.2279&0.2319&0.2276&0.2304&0.2274&0.2297\\
&$\pm\Delta us$&$\pm\Delta us$&$\pm\Delta us$&$\pm\Delta us$&$\pm\Delta us$
&$\pm\Delta us$&$\pm\Delta us$&$\pm\Delta us$
\end{tabular}}
\end{table}
\pagebreak

A glance at table II reveals the following features:

(1) As mentioned above,
both Model I and Model II lead to a significant reduction in $N_\nu$ from 3
(for plausible values of $\eta_u$), in accord with the data. Considering
that $(N_\nu)_{\rm LEP}=2.9841\pm 0.0083$ \cite{27}, we see that Models I and II
would yield nearly the central value of $(N_\nu)_{\rm LEP}$ for
$\eta_u\approx 1/14.5$ and 1/11.8 respectively, and $(N_\nu)_{\rm ESSM}$ would
be within $\pm 1\sigma$ from the central value for
\begin{eqnarray}
\label{eq:etaus}
\eta_u\approx 1/11.5\mbox{ to }1/20\mbox{ (in Model I), and }
\eta_u\approx 1/10\mbox{ to }1/16\mbox{ (in Model II).}
\end{eqnarray}

(2) As discussed in secs. 4, 5 and 6, and as can also be inferred from table 2,
Model II would lead to a reduction in
$(\sigma_{\rm NC}/\sigma_{\rm CC})$ by 0.4 to 0.3\% for $\eta_u=1/11.6$ to
1/13.3, corresponding to $\delta_\mu=0.0040$ to 0.0030. This would (a) reduce
the NuTeV anomaly from a reported $3\sigma$ \cite{NuTeV} to a
$2\sigma$-effect, and simultaneously (b) lead to an increase in the predictions
for $m_H$ and $m_W$ compared to those in the SM (for a fixed $m_{\rm top}$).
Both of these changes appear to be in the right direction, as judged by the
current data, in conjunction with SM theory. By the same token, however,
(with $\delta_\mu\approx 0.4$ to 0.3 \%), Model II would predict (c) a 1 to
1.8$\sigma$ deviation from the present experimental value of the ration
$R_{e/\mu}$ (see sec. 6), and (d) a decrease in the effective values of
$V_{ud}$ and $V_{us}$ (see Appendix A for definitions) by a factor of
$(1-\delta_\mu/2)$ which is in conflict with the currently measured values
and the constraints of CKM-unitarity. (See table II and Appendix A for the
theoretical and observed values of these parameters.)
As expressed in the Appendix, better
judgement in this regard should wait till improved measurements of $V_{us}$
(expected from BNL-E865 and KLOE experiments), together with consistency
checks of different measurements of $V_{ud}$ and reduction of theoretical and
systematic uncertainties in these parameters are in hand \cite{fn51}.

Meanwhile, Model I leads to extremely small values of $\delta_\mu$ (for
$\eta_u$ being in the parameter range that is relevant to measurements of
$N_\nu$ (see eq. \eqref{eq:etaus})). As a result, although it leads to
significant reduction in $N_\nu$ (because $\delta_\tau$ is large), it
practically coincides with the predictions of the SM (see table II and
discussion in sec. 5) as regards (a) NuTeV-measurements, (b) $m_H$ and $m_W$,
(c) $R_{e/\mu}$, and most important (d) $V_{ud}$ and prediction for $V_{us}$
based on CKM unitarity. For example, with $\eta_u=1/18$ (thus
$\delta_\mu=0.00033$), $(N_\nu)_{\rm Model~I}=2.9900$ (in good accord with the
LEP data), but $(\sigma_{\rm NC}^{\rm ESSM}/\sigma_{\rm NC}^{\rm SM})=1-0.03\%$,
and $|V_{us}^{\rm ESSM}|_{\rm unitarity}=0.2276\pm \Delta us$, to be compared
with $|V_{us}^{\rm SM}|_{\rm unitarity}=0.2269\pm \Delta us$.

Thus Model I, while accounting for the LEP data, would be essentially on par
with the SM (especially for $\eta_u\approx 1/28$ to 1/20 (say)), and thereby
free from a large conflict as regards predictions for $V_{ud}$ and $V_{us}$,
that Model II currently seems to have. Model I of ESSM would thus gain
prominence, even over the SM, if (a) the present 2$\sigma$ discrepancy in
$N_\nu$ becomes even a 3 or a 4$\sigma$ effect, but at the same time, (b) NuTeV
anomaly is explained away as a standard model-based QCD and other effects
\cite{Strumia},
{\it and if} (c) improved measurements of $V_{ud}$ and $V_{us}$ remove the
current $2.2\sigma$ discrepancy of the SM in the prediction for $V_{us}$
based on CKM-unitarity. Model II on the other hand could gain prominence
under the following (less likely) set of circumstances: (a) the current
discrepancy in $N_\nu$ stays, (b) NuTeV anomaly survives though at a reduced
level (of 0.3 to 0.4\% (say), see sec. 4), (c) $R_{e/\mu}$ measurement deviates
somewhat from the prediction of the SM (see sec. 6), and most important (d)
$V_{ud}$ and/or $V_{us}$ increase significantly (by say $2\sigma$ each or by
 $2.5\sigma$ in one and  $2\sigma$ in the other) so as to remove the current
discrepancy of Model II in the prediction for $V_{us}$ based on CKM-unitarity.

(3) Both models predict an increase in the tau lifetime compared to the
prediction of the SM, by about 1.0 to 0.38\% in Model I, and 0.45 to 0.16\%
in Model II. These effects are, however, difficult to disentangle at present
from uncertainties in (a) non-perturbative effects for $\tau\rightarrow
\nu_\tau+$ hadrons, and (b) in $\alpha_s$. With a reduction of these
uncertainties, tau lifetime can be an important probe of ESSM.

(4) One general comment is in order. It should be clear from the discussion
above that in a model based on ESSM, if $\nu_{\mu}$-$N'$ mixing is
important
enough to cause noticeable departure from the SM for NuTeV-type measurements
(as in Model II), it would also be important in causing significant departures
from the SM-predictions for (a) $N_\nu$, (b) $m_H$ and $m_W$, (c) $R_{e/\mu}$,
and (d) $|V_{us}|_{\rm unitarity}$, those for the first two being favored and
last two being disfavored
by current data. Generically, one would of course expect $\nu_\mu$-$N'$
mixing in an ESSM-type model to be important, especially if vectorlike leptons
have masses $\lesssim$ 1 TeV (say) (see \cite{FN24}), and thus one would
expect to see
departures from the SM in $N_\nu$ as well as in all the other features listed
above. {\it Model I presents a very interesting and notable exception in this
regard in that it does lead to significant decrease in $N_\nu$, but hardly
affects any of the other features listed above relative to the standard model}.

(5) Last but not least, it is worth recalling here the result of Ref.
\cite{BP(g-2)} that both Models I and II of ESSM can provide a simple
explanation of the indicated anomaly in $(g-2)_\mu$, by utilizing heavy
vectorlike leptons in the loop, and (thus) without requiring sleptons to be
rather light.

\section{A Summary}

Motivations for the ESSM framework have been noted in our earlier papers
\cite{JCP_BPS_BPZ,BP} and are summarized here in
the introduction. We have argued that the  mixing of $\nu_\mu$ and
$\nu_\tau$ with the singlet $N'$ which is naturally expected to arise within
this framework, would generically modify (i) $\nu_\mu$ neutral current
interaction (relevant to the NuTeV-type measurements), (ii) LEP neutrino
counting, (iii) $e$-$\mu$ and $\mu$-$\tau$ universalities in charged current
processes, (iv) predictions for $m_H$ and $m_W$, as well as (v) effective
values of $V_{ud}$ and  $V_{us}$. The degree of modification of these features
depends om the nature of the fermion mass-matrix.

In an earlier work \cite{BPW}, a successful pattern of fermion masses and
mixings (herein called pattern I) has been proposed within a MSSM-based
SO(10)-framework that makes seven predictions all in good accord with
observations, including $V_{cb}\approx 0.04$ and
$\sin^2 2\theta_{\nu_\mu\nu_\tau}\approx 1$. An extension of this framework to
ESSM, preserving the successes of pattern I, has been proposed in a recent
paper, where it is noted that ESSM can provide a simple explanation of the
indicated anomaly in $(g-2)_\mu$ by utilizing contributions from the vectorlike
heavy leptons $(M_{E,E'}\sim$ (300-700) GeV) in the loop.

The main pupose of this paper has been twofold:

(1) First, to bring out new phenomenological possibilities which may arise
within ESSM, in particular to see whether it can be relevant to NuTeV-type
experiments, we have presented here a variant pattern (called pattern II) of
fermion masses and mixings within the SO(10)-framework, and have shown that
the extension of this pattern to
ESSM can account partially for the NuTeV anomaly. The variant pattern is,
of course, interesting in its own right, in that it has the same degree  of
success as pattern I in describing fermion masses and mixings, regardless of
whether it is embedded in ESSM or not.

(2) Second, we have studied here the phenomenological consequences of both
patterns I and II, extended to ESSM (herein called Models I and II). In this
regard, we have argued that both Models I and II lead to a sizable decrease
in LEP neutrino-counting $N_\nu$ (in good acord with the LEP data).
Following the results of Ref. \cite{BP(g-2)}, one can see that both models also
provide a simple explanation of the indicated anomaly in $(g-2)_\mu$. We have
further noted that the two models, interestingly enough, can be clearly
distinguished from each other, however, by other phenomena. In particular,
owing to a relative enhancement of $\nu_\mu$-$N'$ mixing, Model II can
provide (a) a partial explanation of the NuTeV anomaly (as mentioned above),
and simultaneously (b) an increase in the predictions for $m_H$ and $m_W$
compared to the standard model (in accord with the data), (c) small
depatures in $e$-$\mu$ universality in charged current processes (up to 1 to
1.8$\sigma$ deviation from experiments), and (d) small decreases in the
effective values of $V_{ud}$ and $V_{us}$ compared to those in the SM
(which are currently disfavored by the data on grounds of unitarity of the
CKM-matrix). Model I (\cite{BP(g-2),BPW}) on the other hand nearly coincides
with the SM as regards its predictions for these four features: (a)-(d)
(see Table II). (e) Both Models I and II lead to some departures from the
standard model (those for Model I being more prominent) as regards their
predictions for tau lifetime (see table II). Improved experimental and
theoretical studies involving LEP neutrino counting $N_\nu$, $(g-2)_\mu$ and
the five features (a)-(e) listed above can thus not only distinguish between
ESSM versus the standard model, but also between the Models I and II, and
thereby shed light on GUT/string-scale physics.

We stress that the two models I and II, while differing as regards (a)-(d),
share the common feature that both depart from the standard model as regards
their predictions for $N_\nu$ and $(g-2)_\mu$, in good accord with the present
data.

The hallmark of ESSM (irrespective of the NuTeV and LEP neutrino-counting
results) is, of course, the existence of the two complete vectorlike families
$(U,D,N,E)_{L,R}$ and $(U',D',N',E')_{L,R}$ with masses in the range of 200 GeV
to 2 TeV (quarks being heavier than the leptons owing to QCD effects), which
can certainly be tested at the LHC and a future linear collider.

\section*{Acknowledgments}
We have greatly benefitted from discussions and correspondence with W. Marciano,
D. Chowdhury, P. Gambino, G. Isidori, K. McFarland, and A. Strumia.
We wish to thank R. Kellogg for helpful correspondence, and
N.G. Deshpande, D.P. Roy, P. Roy and T. Takeuchi for useful
discussions.   The
work of KSB is supported in part by  DOE Grant \# DE-FG03-98ER-41076,
a grant from the Research Corporation and by
DOE Grant \# DE-FG02-01ER-45684. The work of JCP is supported in part
by DOE Grant \# DE-FG02-96ER-41015.\\

\section*{Appendix A: $V_{ud}$, $V_{us}$ and Unitarity of the CKM Matrix}

First, let us recall the status of $V_{ud}$ and $V_{us}$ and of unitarity
of the $3\times 3$ CKM-matrix in the context of the standard model.
Traditionally, $|V_{ud}|$ is determined by measuring the ratio of $\beta$-decay
and $\mu$-decay rates and using the SM relation: $(G_\beta^{V,A}/G_\mu)_{SM}=
K^{V,A}(V_{ud}^{SM})$, where $K^{V,A}$ denote the ratio of the relevant
matrix elements, including radiative corrections ($V$ and $A$ stand for vector
and axial vector couplings). A very recent analysis leads to a world average
value of $|V_{ud}|$ given by $|V_{ud}|_{\rm SM}^{\rm expt}=0.9739\pm 0.0005$
\cite{Isidori}. This value includes the averages of those based on measurements
 of superallowed nuclear Fermi transitions as well as highly polarized
neutron $\beta$-decay \cite{FN54}. Using this value for $|V_{ud}|$ and
unitarity of the CKM-matrix, one is led to predict
$|V_{us}^{\rm SM}|_{\rm unitarity}=
\sqrt{1-|V_{ud}|^2-|V_{ub}|^2}=0.2269\pm 0.0021$ (where we have used
$|V_{ub}|_{\rm SM}^{\rm expt}=0.0038\pm 0.0008$; $|V_{ub}|$ has, of course,
a negligible effect on this prediction). The direct determination of
 $V_{us}$, obtained from $K_{l3}$ decay rates, however, yields:
 $|V_{us}|_{\rm SM}^{\rm expt}=0.2196\pm 0.0026$ \cite{ref55,Isidori} which is
about 2.2$\sigma$ {\it below} the prediction based on CKM-unitarity. It has
been expressed in Ref. \cite{Isidori} that such a discrepancy may have its
origin in an underestimate of theoretical and systematic errors. An improved
determination of $|V_{us}|$ is expected from high-statistics measurements
of $K_{l3}$ decay rates at the BNL-E865 \cite{ref56} and KLOE experiments
\cite{ref57}. It is worth mentioning here that preliminary results of the
BNL-E865 experiment indicate that $|V_{us}|=0.2278\pm 0.0029$
\cite{ref56,Isidori}. Such a value, if it holds up, would of course be in
good accord with the constraints of unitarity of the CKM-matrix.
It thus seems that a better assessment of the extent to which
CKM-unitarity is respected, {\it in the context of even the standard model},
should
wait untill improved values of $|V_{us}|$ and $|V_{ud}|$, together with
reduction of theoretical and systematic uncertainties are available.

Turning to ESSM, one would expect, in principle, two modifications here
pertaining to quark-mixings: (i) Even in the absence of SU(2)$_L$-singlet
quarks $(U',D')$, the $3\times 3$ CKM-matrix would extend to a
$4\times 4$-matrix, to be denoted by $\tilde V_{ij}$, which describes mixings
among the four SU(2)$_L$-doublets
$(q_L^{e,\mu,\tau}\mbox{ and }Q_L=(U_L,D_L))$. The
CKM-unitarity constraint is then extended to satisfy conditions such as
$\tilde V_{ud}^2+\tilde V_{us}^2+\tilde V_{ub}^2+\tilde V_{uD}^2=1$. (ii)
Allowing for the $u-U'$-mixing (one can safely neglect $d-D'$-mixing, because
$\theta_{dD'}/\theta_{uU'}\sim (\kappa_d/\kappa_u)\sim m_b/m_t\ll 1$), each
of the tilde-entries involving the up-quark, which together satisfy the
unitarity-constraint as above, would be reduced by $(1-\delta_u/2)$, so that
$V_{uj}^{\rm ESSM}=\tilde V_{uj}(1-\delta_u/2)$, where $j=d,s,b,D$ and
$\delta_u=\theta^2_{uU'}$.

In practice, however, each of these two modifications turns out to be
insignificant, especially for mixing elements involving the up-quark. This is
primarily because one can always transform from the gauge to the hat-basis
(see discussions in sec. 2 and Appendix B) in which the first family gets
decoupled from the vectorlike families. We estimate \cite{ref58}:
$\theta_{u_LU'_L}\lesssim (1/3)(10^{-2})$ and
$\theta_{u_LU_L}\lesssim 10^{-3}$. So, $\delta_u=\theta^2_{uU'}\leq 10^{-5}$,
and $|V_{uD}|=|\theta_{uU'}-\theta_{dD}|\leq 10^{-3}$. Thus, even in ESSM,
we would expect that the unitarity-constraint of the $3\times 3$ CKM-matrix,
involving especially the up-quark -- i.e., the condition
$( V_{ud}^2+V_{us}^2+ V_{ub}^2)_{\rm ESSM}=1$ -- should
be maintained, barring  corrections given by $ V_{uD}^2\leq 10^{-6}$
and $\tilde V_{ud}^2\delta_u\leq 10^{-5}$. While these corrections might have
been more important in a more general context, we can safely ignore them
henceforth in the context of the model under discussion.

There is, however, a modification in the effective value of  $V_{ud}$ in
ESSM compared to its standard model-based experimental value, because the
latter is extracted by utilizing the ratio of $\beta$-decay and $\mu$-decay
rates, and there is a modification in the theoretical expression for $G_\mu$ in
ESSM compared to that in the SM (see Eq. \eqref{eq:Gmu}). The same modification
would apply also to $V_{us}$ which is determined  from the ratio of $K_{l3}$
and $\mu$-decay rates, and likewise for $V_{ub}$.

Writing $G_\beta^{\rm ESSM}\propto (g^2/8m^2_W)V_{ud}^{\rm ESSM}$, and
$G_\mu^{\rm ESSM}\propto (g^2/8m^2_W)(1-\delta_\mu/2)$ (where we have
suppressed radiative corrections which are partly common to $\beta$-decay and
$\mu$-decay, see Eq. \eqref{eq:Gmu}), we obtain:
$(G_\beta/G_\mu)_{\rm ESSM}=V_{ud}^{\rm ESSM}/(1-\delta_\mu/2)=
(V_{ud})_{\rm SM}^{\rm expt}$; and thus
\begin{eqnarray}
\label{eq:A1}
|V_{ud}|_{\rm ESSM}^{\rm expt}=|V_{ud}|_{\rm SM}^{\rm expt}(1-\delta_\mu/2),
\end{eqnarray}
and likewise $|V_{us}|_{\rm ESSM}^{\rm expt}=
|V_{us}|_{\rm SM}^{\rm expt}(1-\delta_\mu/2)$, and
$|V_{ub}|_{\rm ESSM}^{\rm expt}=|V_{ub}|_{\rm SM}^{\rm expt}(1-\delta_\mu/2)$.
To see the effect of the reduction factor $(1-\delta_\mu/2)$ on the unitarity
constraint, we write:
\begin{eqnarray}
\label{eq:A2}
V_{ud}^{\rm ESSM}=V^0_{ud}(1-\delta_\mu/2)\pm\Delta,
\end{eqnarray}
where $V^0_{ud}$ denotes the central value of $|V_{ud}|_{\rm SM}^{\rm expt}$,
and $\pm\Delta$ denotes the error in it (with $\Delta >0$). Using the
unitarity condition for the first row of the CKM-matrix (which holds for
ESSM, barring corrections $\lesssim 10^{-5}$), we would predict:
\begin{eqnarray}
\label{eq:A3}
|V_{us}|_{\rm unit}^{\rm ESSM}=\sqrt{1-(V_{ud}^2+V_{ub}^2)_{\rm ESSM}}=
|V^0_{ud}|_{\rm unit}+\frac{(V^0_{ud})^2(\delta_\mu/2)}{(V^0_{us})_{\rm unit}}
\pm \frac{V^0_{ud}}{(V^0_{us})_{\rm unit}}\Delta,
\end{eqnarray}
where $|V^0_{us}|_{\rm unit}$ is the central value for $V_{us}$ that is
predicted by using the SM-based central value of
$|V_{ud}|_{\rm SM}^{\rm expt}$, together with CKM-unitarity. Thus
\begin{eqnarray}
\label{eq:A4}
|V^0_{us}|_{\rm unit}\equiv\sqrt{1-(V_{ud}^0)^2-(V_{ub}^0)^2}.
\end{eqnarray}
In writing Eq. \eqref{eq:A3}, we have ignored the error in $V_{ub}$ since even
$V_{ub}^0$ is immaterial.

If one uses the world-average value of $|V_{ud}|_{\rm SM}^{\rm expt}=0.9739
\pm 0.0005$ \cite{Isidori}, then $|V^0_{ud}|=0.9739$ and $\Delta=0.0005$, while
$|V_{ub}^0|=0.0038$. One then predicts, within the SM, $|V^0_{us}|_{\rm unit}=
0.2269$, and thus (using Eq. \eqref{eq:A3}),
\begin{eqnarray}
\label{eq:A5}
|V_{us}|_{\rm unit}^{\rm ESSM}=0.2269+4.179(\delta_\mu/2)\pm \Delta us,
\end{eqnarray}
where $\Delta us \equiv 4.29(\Delta)=0.0021$
(for $\Delta=0.0005$). We will use Eq. \eqref{eq:A5}
in computing the entries in table II of sec. 7.\\

\section*{Appendix B: Inclusion of the First Family: Relevance To Solar
Neutrino Oscillation?}

As remarked in sec. 2, the matrices $(X,X',Y\mbox{ and }Y')$, which denote the
mixings of the three chiral with the vectorlike families in the gauge basis, can
always be transformed to their hat-forms $(\hat X,\hat X',\hat Y\mbox{ and }
\hat Y')$ so that one family is left exactly massless, which can then be
identified with the electron family. Thus, the masses of the $(u,d,e,(\nu_e)_D)$
must arise {\it entirely} through the direct mass terms of order few MeV
(that enter into the $0_{3\times 3}$ block of Eq. \eqref{eq:MD}). But the
mixings of the electron family with the $\mu$ and the $\tau$-families can arise
not only from the direct mass terms but also from the transformation from the
gauge to the hat  basis (i.e., $X\rightarrow \hat X$,  $X'\rightarrow \hat X'$,
etc.). As we now discuss, the mixings from this latter source can in
principle be important especially for the first family, under the circumstances
which we note below. To illustrate this possibility, we will consider only
Model II. Qualitatively similar results will apply to Model I as well.

To include the first family $({\bf 16}_1)$, we start in the gauge basis and add
 the following effective terms to the superpotential for Model II, given in Eq.
\eqref{eq:WrmYuk}:
\begin{eqnarray}
\label{eq:B1}
W_{\rm Yuk} &=& h_{1V} {\bf 16}_1 {\bf 16}_V {\bf 10}_H + h'_{1V}
{\bf 16}_1 {\bf 16}_V {\bf 16}_H{\bf 16}_H/M+\tilde h_{1V} {\bf 16}_1
{\bf 16}_V {\bf 10}_H{\bf 45}_H/M.
\end{eqnarray}
The vectors $X_f$ and $X'_f$ of Eq. \eqref{eq:MYuk} will thus have the form:
\begin{eqnarray}
\label{eq:B2}
X_f=(x_1,x_2,x_3)_f^T,\quad X'_f=(x'_1,x'_2,x'_3)_f.
\end{eqnarray}
At the GUT scale, the vectors $X'_f$ (defined in the gauge basis) are given by:
\begin{eqnarray}
\label{eq:B3}
X'_u=[2(\sigma'+\epsilon'),2\epsilon,1]x'_u;\quad \quad
X'_d=\left[ \frac{2(\eta'+\epsilon')}{1+\xi},
 \frac{2(\eta+\epsilon)}{1+\xi},1\right]x'_d\nonumber\\
X'_\nu=[2(\sigma'-3\epsilon'),-6\epsilon,1]x'_u;\quad
X'_l=\left[ \frac{2(\eta'-3\epsilon')}{1+\xi},
 \frac{2(\eta-3\epsilon)}{1+\xi},1\right]x'_d\
\end{eqnarray}
The vectors $X_f$ are obtained from Eq. \eqref{eq:B3} by the replacement
$(\epsilon,\epsilon')\rightarrow -(\epsilon,\epsilon')$. Here $\sigma'$ and
$\epsilon'$ are proportional to $h_{1V}$ and $\tilde h_{1V}$ respectively,
while $\eta'\equiv \hat\eta'+\sigma$, with $\hat\eta'$ being proportional
to  $h'_{1V}$. The parameters $\eta$, $\epsilon$ and $\xi$ have been defined
before (see sec. 3).

As mentioned above, the mixing angle $V_{ij}$ involving the first family get
contributions both from the direct mass terms, as well as from Eq. \eqref{eq:B1}
through transformations of $X'_f$, etc. from the gauge to the hat basis. Let us
denote the two contributions to $V_{ij}$ by $V_{ij}^{\rm dir}$ and $V'_{ij}$
respectively; thus (for $i\neq j$, with small mixing angles $\leq 1/3$ (say))
$V_{ij}\approx V_{ij}^{\rm dir}+V'_{ij}$. We can bring
Eq. \eqref{eq:L} into the hat basis, where the electron family decouples, by
making a rotation in the 1-2 family-space. For example, a rotation in the
$u_L-c_L$ quark sector by an angle
$\tan \theta'_{uc}=(\sigma'+\epsilon')/\epsilon$ will bring $X'_u$ to the
hat-form $[0,2\sqrt{\epsilon^2+(\epsilon'+\sigma')^2},1]x'_u$, etc. Now we
can see that even with a very mild  hierarchy between the entries for the first
and the second families (in the gauge basis), the discussion of the 2-3 sector
carried out in the previous section will remain essentially unchanged. For
instance, even for a rather large value of the ratio
$|\epsilon'+\sigma'|/|\epsilon|\approx 1/2$ (say), the correction terms
relevant to the 2-3 sector are only of order
$(1/2)(\epsilon'+\sigma')^2/\epsilon^2\approx 12\%$.

At the same time, such a large ratio is fully compatible with the smallness
of the up-quark mass because the $(\epsilon'+\sigma')$-entry contributes zero
to $m_{\rm up}$ by the rank argument. This is the new feature of ESSM. It
provides an entirely {\it new source} for mixing of $(u,d,e,\nu_e)$ with
the corresponding members of the second and the third families, which in
principle could be relatively large without conflicting with the smallness of
their masses. Of course, if flavor symmetries, distinguishing between the
three families, dictate a large hierarchy already in the gauge basis, so
that $x'_1\ll x'_2 < x'_3$ etc., such mixings denoted by $V'_{ij}$ would be
small. In this appendix, for purposes of illustration, however, we would
allow for possible departures from such a hierarchical structure.

Including a rotation in the $d_L$-$s_L$ sector (analogous to the $u_L$-$c_L$
rotation), we obtain the following contributions (from the transformations
of the $X'_{u,d}$ matrices) to the CKM mixings:
\begin{eqnarray}
\label{eq:B4}
V'_{us}=\frac{\sigma'+\epsilon'}{\epsilon}-\frac{\eta'+\epsilon'}
{\eta+\epsilon},\quad V'_{ub}\approx V'_{us}\theta_{sb},
\end{eqnarray}
where $\theta_{sb}$ is the contribution to $V_{cb}$ arising from the
diagonalization of the mass matrix $(M_d)$ from the down sector
(see Eq. \eqref{eq:4matrices}). Thus, $\theta_{sb}=\eta+\epsilon=0.0386$
(using values for $\eta$ and $\epsilon$ given in Eq. \eqref{eq:values}). It is
clear from Eq. \eqref{eq:B4} that $V_{us}$ and $V_{ub}$ cannot be generated
entirely or primarily through the mixings of the chiral with the vectorlike
families. Because if $V_{us}\approx V'_{us}$ and $V_{ub}\approx V'_{ub}$,
then Eq. \eqref{eq:B4} would yield $V_{ub}\approx (0.22)(0.0386)\approx 0.008$,
which is a factor of two larger that the experimental value. The direct mass
terms (entering into the $3\times 3$ block of Eq. \eqref{eq:MD}), which should
be of order $m_d\sim$ few MeV, are expected to contribute significantly to
these mixings. We would expect $V_{ub}^{\rm dir}\sim{\cal O}(m_d)/m_b\sim
{\cal O}(0.0015)$ which is of the same order as $V_{ub}$, and similarly
$V_{us}^{\rm dir}\sim{\cal O}(m_d)/m_s\sim 0.06$.

Turning to the neutrino sector, the $\nu_e$-$\nu_\mu$ oscillation angle induced
by the transformations of $X'_\nu$ and $X'_l$ to their hat-forms is given by:
\begin{eqnarray}
\label{eq:B5}
\theta^{\prime {\rm osc}}_{\nu_e\nu_\mu}\approx\frac{x'_{1\nu}}{x'_{2\nu}}-
\frac{x'_{1l}}{x'_{2l}}=\frac{\sigma'-3\epsilon'}{-3\epsilon}
-\frac{\eta'-3\epsilon'}{\eta-3\epsilon}.
\end{eqnarray}
Combining with the expression for $V'_{us}$ in Eq. \eqref{eq:B4}, we obtain:
\begin{eqnarray}
\label{eq:B6}
\theta^{\prime {\rm osc}}_{\nu_e\nu_\mu}\approx\frac{V'_{us}(1+\eta/\epsilon)}
{3-\eta/\epsilon}\left[\frac{(\eta/\epsilon)(1-\sigma'/(3\epsilon'))+
(\sigma'/\epsilon'-\eta'/\epsilon')}{(\eta/\epsilon)(1+\sigma'/\epsilon')+
(\sigma'/\epsilon'-\eta'/\epsilon')}\right].
\end{eqnarray}
It is worth asking whether Eq. \eqref{eq:B6} can consistently lead to a large
oscillation angle for solar neutrinos ($\sim 0.5$, as suggested by current data
\cite{solarLMA}). One way to obtain a large value for
$\theta^{\prime {\rm osc}}_{\nu_e\nu_\mu}$ is to maximize the numerator and minimize the
denominator in the square bracket of Eq. \eqref{eq:B6}. [Note $\eta$ and
$\epsilon$ are fixed in the model (here we are considering Model II presented
in sec. 3) from our considerations of fermion masses
(see Eq. \eqref{eq:values})]. To this end, parametrizing
$\sigma'/\epsilon'\equiv -1+r_1$ and $\eta'/\epsilon'\equiv -1+r_2$
(with the presumption that $r_1,r_2\ll 1$), we obtain:
\begin{eqnarray}
\label{eq:B7}
\theta^{\prime {\rm osc}}_{\nu_e\nu_\mu}\approx V'_{us}\left[\frac{0.382}
{0.772\, r_1+r_2}\right],\quad  V'_{us}\approx (\epsilon'/\epsilon)\left[
\frac{0.772\, r_1+r_2}{0.772}\right],
\end{eqnarray}
where we have inserted $\eta=0.0886$ and $\epsilon=-0.05$
(see Eq. \eqref{eq:values}). This in turn yields
$\theta^{\prime {\rm osc}}_{\nu_e\nu_\mu}\approx (0.382/0.772)(\epsilon'/\epsilon)
\approx (0.50)(\epsilon'/\epsilon)$. Thus, keeping
$|\epsilon'/\epsilon|\leq 1/2$ (this is to preserve the predictivity and the
success of the 2-3 sector discussed in sec. 3), we see that the new source
of mixing for the first family available in ESSM can provide a sizable value
for the $\nu_e$-$\nu_\mu$ oscillation angle
$\theta^{\prime {\rm osc}}_{\nu_e\nu_\mu}\leq 0.25$, but not quite as big as the large
angle ($\sim 0.5$) suggested by the data \cite{solarLMA}. The $\nu_e$-$\nu_\mu$
mixing can of course receive contributions of the desired magnitude from
other sources consistent with the SO(10)/G(224)-symmetry (see e.g., Ref.
\cite{JCP_Erice}). We note that given $\nu_e$-$\nu_\mu$ mixing from the source
as above (Eq. \eqref{eq:B6}), and $\nu_\mu$-$N'$ mixing discussed in sec. 4,
we will have $\nu_e$-$N'$ mixing given by
\begin{eqnarray}
\label{eq:B8}
\theta_{\nu_e N'}\approx (\theta^{\prime {\rm osc}}_{\nu_e\nu_\mu})
(\theta_{\nu_\mu N'}),
\end{eqnarray}
where $\theta_{\nu_\mu N'}=\sqrt{\delta_\mu}=|\eta_u(c_Lp'_\nu-s_L)|$
(see sec. 4). The effects of $\nu_e$-$N'$ mixing are reflected in the relevant
expressions in sec. 4.

}

\begin{thebibliography}{30}
\bibitem{FN4} It is pressumed (see e. g., Ref. \cite{JCP_BPS_BPZ} and
\cite{BP}) that the mass-term $M_V$ of the vectorlike families is protected
by some local ``generalized flavor" or discrete symmetries (presumably of
string origin), so that it is of order one TeV, rather that the GUT-scale,
just like the $\mu$-term of MSSM. The same set of symmetries are also assumed
to ensure that the three chiral families receive their masses primarily
through their mixings with the two vectorlike families rather than through
direct couplings with each other (see remarks later).
\bibitem{FN8} This mechanism for inter-family mass hierarchy was also noted
by S. Barr [Phys. Rev. {\bf D42}, 3150 (1990)]. In this work, the vectorlike
families were
assumed to have GUT-scale masses, which would not be relevant to present
considerations.
\bibitem{FN20} The right-handed neutrinos of three chiral families
($\nu_R^e$, $\nu_R^\mu$ and  $\nu_R^\tau$) acquire superheavy Majorana
masses, which, following the familiar see-saw mechanism, leave the
left-handed neutrinos light. Since RH neutrinos are superheavy, the
mixing of LH neutrinos $\nu_L^{e,\mu,\tau}$ with $(N,N')_L$, obtained from
Eq. \eqref{eq:MD}, would not be affected by their presence.
\bibitem{FN25} The electron family
is essentially massless, barring small entries ($\leq$ a few MeV) in the
direct $0_{3 \times 3}$ subsector of Eqs. \eqref{eq:MYuk} and
\eqref{eq:MD}.
\bibitem{FN27} Since corrections to see-saw diagonalization for the
$\mu-\tau$ sector are not small, especially for $m_\mu^0/m_\tau^0$, the
results quoted are based on a more accurate numerical diagonalization.
\bibitem{FN26} The corresponding values of the $p$-parameters for Model I
of ESSM \cite{BP(g-2)} are given in sec. 7, where the predictions of
Models I and II are compared.
\bibitem{FN26_} Note, despite QCD effects, the parameters $\eta_u$ for the
neutrinos and for up-quarks are essentially the same.
\bibitem{FN37} This difference $\Delta r'_W$ arises because of the difference
between radiative corrections involving quark-lines (as in
$\nu_\mu N\rightarrow \nu X$) and purely leptonic lines (as in $\mu$-decay),
through vertex and box graphs. This difference is, however, the same for ESSM
as for the SM.
\bibitem{newrad} We are ignoring here the radiative corrections due to
SUSY and due to ESSM, which seem to be rather small for our purposes.
For example, the correction to the $\rho$ parameter from the vectorlike
quarks $U$ and $D$  of ESSM can be estimated to be
$\Delta \rho^{\rm ESSM} \approx (\Delta \rho^{\rm top}/3)
[(m_U^2 - m_D^2)^2]/(m_t^2 m_U^2) \approx (\eta_u^4/3) (m_U^2/m_t^2)
\Delta \rho^{\rm top} \approx 3 \times 10^{-6}$, where we used $m_U
\simeq m_D \sim 500$ GeV
and $\eta_u \sim 1/10$ for the numerical estimate.
The estimate is similar for $\Delta \rho$
correction from SUSY particles, if the masses of the SUSY particle are of
order 500 GeV.
\bibitem{FN44} Constraint on increase in $m_W$ from measurements of
$\sin^2\theta_W$ is noted below (see \cite{FN47}).
\bibitem{langacker2} Using Ref. \cite{erler}, one would obtain a somewhat
higher value of $m_H \approx 83$ GeV, which is still considerably lower
than the ESSM value.
\bibitem{FN46} Change in $m_t$ from its central value (174.3 GeV) by $\pm 5.1$
GeV would induce a shift in $m_W$ by $\pm 32$ MeV, as well as a shift in the
prediction for $m_H$ (see Ref. \cite{degrassi} for details). For the sake of
concretenes, columns (c) and (d) in Table I are obtained by keeping the
difference in the predictions of $m_W$ for the two cases (SM versus ESSM)
fixed at 20 MeV.
\bibitem{FN47} As noted above,
increase in $m_W$ by say 20-40 MeV (compared to the SM) would typically be in
better accord with the data. This would, however, lead to a decrease in
the prediction for the on-shell value of $\sin^2\theta_W=1-m^2_W/m^2_Z$ by
$\delta(s^2_W)=-0.00039$(1 to 2) for $\delta m_W=$(20-40) MeV. A decrease
by (say) 0.00039 is comparable to the experimental
error in $\sin^2\theta_W$ obtained from Standard Model fit by LEPEWWG
which gives $\sin^2\theta_W=0.2227\pm 0.00037$ \cite{27}, and thus quite
acceptable. By itself, such a decrease would of course have negligible
bearing on the anomaly in the NuTeV
result, which yields $(\sin^2\theta_W)_{\rm on~shell}
=0.2277\pm 0.0013\pm 0.0009$ (ignoring
negligible corrections through $m_t$ and $m_H$), if the overall strength of
the NC amplitude is kept at the SM value \cite{NuTeV}.
\bibitem{fn51} In particular, if the true value of $|V_{us}|$ turns out to be
higher than the present value of $0.2196\pm 0.0026$ \cite{Isidori,ref55}; for
instance, if the central value of $|V_{us}|$ is closer to that of the
preliminary value
of $0.2278\pm 0.0029$ of the E865 experiment \cite{Isidori,ref56}, noted in
Appendix A, the conflict in this regard with respect to Model II would be much
milder (about 1.2$\sigma$ for $\eta_u=1/15$).
\bibitem{ref58} As discussed in sec. 2 and in particular in Appendix B, as one
transforms from the gauge to the hat basis in two steps: (a) the $Y^T$ and
$Y'$ matrices are transformed into the forms $\propto (0,0,1)$, and next
(b) the resulting $X^T_f$ and $X'_f$-matrices are transformed into forms
$\propto (0,p_f,1)$ and $\propto (0,p'_f,1)$ respectively; the first
family gets decoupled from the vectorlike families. In this process, one
introduces $u_L$-$c_L$ mixing in step (b) given by $\theta_{u_Lc_L}\sim
(x'_1/x'_2)_{\rm up}\lesssim 1/3$ (say, see Appendix B).
We thus estimate that $\theta_{u_LU'_L}\approx
(\theta_{u_Lc_L})(\theta_{c_LU'_L})\lesssim (1/3)(p'_u\eta_u)\lesssim (1/3)
(10^{-2})$ (where we have put $\eta_u\leq 1/10$; $p'_u\approx 0.03$ (for Model
I), and $|p'_u|\approx 0.1$ (for Model II)). Furthermore, using a form for $M_u$
analogous to Eq. \eqref{eq:MD}, we estimate that $\theta_{u_LU_L}\approx
(\theta_{u_Lc_L})(\theta_{c_LU'_L})(\theta_{U'_LU_L})\approx
(\lesssim 1/3)(p'_u\eta_u)(\sim 1/3\mbox{ to }1/2)\lesssim (10^{-3})$, as
mentioned in the text.
\end{thebibliography}
\end{document}